\pdfoutput=1 
\documentclass[aps,prb,twocolumn,showpacs,amsmath,amssymb,superscriptaddress,10pt,nofootinbib]{revtex4-2}

\usepackage[utf8]{inputenc}
\usepackage[T1]{fontenc}

\usepackage[pdftex]{graphicx}
\usepackage[colorlinks,linkcolor=blue,citecolor=blue,urlcolor=blue]{hyperref}

\usepackage{xcolor} 
\usepackage{ulem} 
\usepackage{dsfont}
\usepackage[multidot]{grffile}

\usepackage{physics}

\newcommand{\mat}{\mathbf}

\newcommand{\ii}{\mathrm{i}}
\newcommand{\ee}{\mathrm{e}}
\newcommand{\kB}{k_\mathrm{B}}

\newcommand{\bS}{\begin{subequations}}
\newcommand{\eS}{\end{subequations}}

\newcommand{\aL}{\mathrm L}
\newcommand{\aR}{\mathrm R}

\newcommand{\vs}{\vec{s}}
\newcommand{\TM}{\mat T}
\newcommand{\TMU}{\mat U}
\newcommand{\TMV}{\mat V}
\newcommand{\dtU}{\delta_U}

\newcommand{\Zp}{\check Z} 

\newcommand{\DC}{\tilde\Sigma^{C}}
\newcommand{\DO}{\tilde\Sigma^{O}}

\newcommand{\tm}{t_m} 
\newcommand{\tb}{t_\mathrm b} 
\newcommand{\ta}{t_\mathrm a}
\newcommand{\Lb}{L_\mathrm b}
\newcommand{\La}{L_\mathrm a}
\newcommand{\mm}{\tilde m}

\DeclareMathOperator{\diag}{\mathrm{diag}}

\begin{document}
 \title{Transfer-matrix summation of path integrals for transport through nanostructures}

\author{Simon Mundinar}
\affiliation{Theoretische Physik, Universität Duisburg-Essen and CENIDE, 47048 Duisburg, Germany}

\author{Alexander Hahn}
\affiliation{Theoretische Physik, Universität Duisburg-Essen and CENIDE, 47048 Duisburg, Germany}

\author{Jürgen König}
\affiliation{Theoretische Physik, Universität Duisburg-Essen and CENIDE, 47048 Duisburg, Germany}

\author{Alfred Hucht}
\affiliation{Theoretische Physik, Universität Duisburg-Essen and CENIDE, 47048 Duisburg, Germany}

\date{\today}

\begin{abstract}
On the basis of the method of iterative summation of path integrals (ISPI), we develop a numerically exact transfer-matrix method to describe the nonequilibrium properties of interacting quantum-dot systems. 
For this, we map the ISPI scheme to a transfer-matrix approach, which is more accessible to physical interpretation, allows for a more transparent formulation of the theory, and substantially improves the efficiency. 
In particular, the stationary limit is directly implemented, without the need of extrapolation.
The resulting new method, referred to as ``transfer-matrix summation of path integrals'' (TraSPI), is then applied to resonant electronic transport through a single-level quantum dot.
\end{abstract}

\maketitle

\section{Introduction}
Quantum-dot systems have been well studied both experimentally and theoretically for over thirty years. 
Their optical properties, namely the quantum size effect, make them useful for commercial applications within liquid crystal displays \cite{Nanotech_2014}.
The tunability of their electrical properties allows to control single electrons \cite{Splettstoesser_2017} and gives rise to a number of effects such as Coulomb blockade, the Kondo effect, tunnel magnetoresistance or Andreev bound states \cite{Pekola_2013}, on which versatile electronic and spintronic quantum-dot devices such as a single-electron transistor \cite{Wang_2010, Devoret_2000}, a quantum-dot spin valve \cite{Sahoo_2005, Crisan_2016} or a Cooper-pair splitter \cite{Schindele_2014, Tan_2015, Borzenets_2016} are based. 

In order to study resonant transport in interacting quantum-dot setups, a method called ``iterative summation of path integrals'' (ISPI) was developed \cite{Weiss_2008, Weiss_2013, Mundinar_2019}. 
ISPI is an exact-enumeration method that is based on the systematic truncation of correlations decaying exponentially in time. 
The ISPI method is well suited to study quantum-dot systems at finite temperature, including both equilibrium and nonequilibrium, in the regime in which various energy scales, e.g., associated with Coulomb interaction, temperature, or transport voltage, are of the same order of magnitude and, therefore, lack a clear separation.
The ISPI scheme was first introduced to discuss nonequilibrium transport through the Anderson model \cite{Weiss_2008, Weiss_2013}. 
Applying the method to the Anderson-Holstein model, where the quantum dot is coupled to a phonon mode, demonstrated the impact on the Franck-Condon blockade, when entering the quantum-coherent regime \cite{Huetzen_2012}. 
Recently, the ISPI method was applied to quantum-dot spin valves, demonstrating the importance of resonant effects in the tunnel-magnetoresistance as well as unveiling interaction-induced current asymmetries caused by an interaction-induced exchange field \cite{Mundinar_2019, Mundinar_2020}.
A comparison of ISPI with other methods was done in \cite{Eckel_2010}.

The purpose of this work is to develop the ISPI scheme further. 
We show that the necessary truncation of correlations motivates a mapping of ISPI to a transfer-matrix approach, which by construction is formulated in the stationary limit, such that extrapolation of finite-time results is not needed anymore.
We develop the theoretical corner stones of this new method, referred to as ``transfer-matrix summation of path integrals'' (TraSPI). 
In order to keep the discussion transparent, we exemplify the method for a system with relatively few degrees of freedom, namely the Anderson model describing a single-level quantum dot coupled to two normal metal leads. 
We, however, emphasize that all new concepts discussed in this work are not limited to this simple model, but can easily be transferred to other, more intricate setups, such as hybrid quantum-dot systems involving superconducting and/or ferromagnetic leads \cite{Sahoo_2005, Crisan_2016,Schindele_2014, Tan_2015, Borzenets_2016}, or quantum-dot Aharonov-Bohm interferometers \cite{Aikawa_2004,Ihn_2007,Edlbauer_2017}, just to name a few.

Single-electron transistors that utilize a quantum dot as an island have gathered a lot of attention throughout the years and are still under heavy investigation, both experimentally and theoretically. 
To mention just a few recent examples, a single-electron transistor consisting of a quantum dot and normal metal leads was realized experimentally to demonstrate that shot noise in a single-electron transistor can be reduced significantly via feedback control, which should allow the construction of efficient, nanoscale thermoelectric devices \cite{Wagner_2017}. 
Furthermore, if the quantum dot is periodically driven via a gate voltage, it is possible to accurately control the dot's emission time statistics \cite{Brange_2021}. 
For a system, in which the quantum dot is coupled to a single lead only, the quantum dot can be driven out of equilibrium via a plunger gate voltage, which allows the measurement of the free energy of a confined electron in order to study thermodynamics on the microscopic level \cite{Hofmann_2016}. 
For a superconducting single-electron transistor, an attractive interaction was found that survives even far beyond the superconducting regime \cite{Guenevere_2017, Cheng_2015}. 
Employing the method of full-counting statistics for a negative-$U$ Anderson model, it was shown that this phenomenon is robust, even for fast spin relaxation \cite{Kleinherbers_2018}.

On a theoretical basis, different approaches are used and actively developed to study different parameter regimes of quantum-dot systems. 
The method of ``Dynamical Mean Field Theory'' is advanced and combined with other methods, such as functional renormalization group theory, to increase the predictive power of the method, even for strong and nonlocal electron correlations \cite{Rohringer_2018}. 
Perturbation theory in the tunnel coupling strength $\Gamma$ within a master-equation approach, has proven highly useful in the description of quantum-dot systems. 
This method is developed further in different directions, e.g., by introducing SU($N$)-invariant kinetic equations to effectively study multilevel quantum dots \cite{Maurer_2020} or by improving on the commonly used rotating-wave approximation, leading to a so-called coherent approximation \cite{Kleinherbers_2020}. 
While perturbative methods often allow for at least a qualitative description, nonperturbative effects are, by construction, beyond their scope. 
To cover them, numerically exact methods are in high demand.
Several approaches are known to tackle this problem. 
Quantum Monte Carlo simulations were advanced to reach the stationary regime for systems in nonequilibrium \cite{Profumo_2015, Bertrand_2019}. 
Different flavors of renormalization group theory (RG) have been applied to quantum-dot systems almost since the inception of their theoretical discussion \cite{Anderson_1970}. 
Since then significant advances have been made to the formalism: A combination of numerical RG and time-dependent density-matrix RG allows to discuss the nonequilibrium steady state transport properties of quantum-dot systems \cite{Schwarz_2018}, while within functional RG it was possible to approximate the flow of the Luttinger-Ward functional while maintaining conservation laws \cite{Rentrop_2016}. 
Finally, it was shown that density functional theory is able to study out-of-equilibrium transport theories, even in strongly correlated systems, such as the Anderson model \cite{Kurth_2016}, while a quasiparticle Fermi-liquid theory can be used to work within the low-energy limit of such systems \cite{Mora_2015}.

The article is structured as follows. 
In Sec.~\ref{sec:Model} we introduce the Anderson model's Hamiltonian and derive the path-integral formulation of the generating function, as well as its discrete counterpart. 
We demonstrate how interactions are decoupled via a discrete Hubbard-Stratonovich transformation and then solve the path integral.  In Sec.~\ref{sec:Method} we discuss the main ideas behind the ISPI and TraSPI schemes, namely the systematic truncation of exponentially decaying correlations after a memory time $t_K$, as well as the subsequent block factorization of the Keldysh partition function.
We then demonstrate how the ISPI scheme can be mapped to a transfer-matrix approach, and discuss the main benefits of this new formulation in Sec.~\ref{sec:TMImplementation}. 
We finish this section with a description of the two-step extrapolation procedure that eliminates both the introduced discretization error as well as the truncation error, and therefore ensures that the results obtained via the TraSPI scheme are numerically exact within the accuracy of the eigensystem calculation. 
In Sec.~\ref{sec:Results} we discuss the results for the Anderson model, obtained via TraSPI. 
First, we discuss current based observables, like the current and the conductance, and after that the dot's occupation number. 
Finally, we conclude in Sec.~\ref{sec:Conclusion}. 

\section{Model} \label{sec:Model}
We write the well-known Hamiltonian for an interacting, single-level quantum dot that is tunnel-coupled to two metallic leads in the form \cite{Anderson_1961, Weiss_2008} (we set $\hbar=1$ throughout this work)
\begin{align}\label{eq:Hamiltonian}
	\mathcal{H} = & \sum_\sigma \epsilon_{0,\sigma} \hat n_\sigma - \frac{U}{2} \left(  \hat n_\uparrow - \hat n_\downarrow  \right)^2 \nonumber\\
    & + \sum_{\alpha \vb{k} \sigma} \epsilon_{\alpha \vb{k}} \hat c^\dagger_{\alpha\vb{k}\sigma} \hat c_{\alpha\vb{k}\sigma}
	+ \sum_{\alpha \vb{k} \sigma} \big(t_\alpha \hat c^\dagger_{\alpha\vb{k} \sigma} \hat d_{\sigma} + \text{h.c.}\big).
\end{align}
The on-site occupation-number operator is given by $\hat n_\sigma = \hat d^\dagger_\sigma \hat d_\sigma$, where $\hat d^\dagger_\sigma$ and $\hat d_\sigma$ create or annihilate an electron on the quantum dot with spin $\sigma = \,\uparrow,\downarrow$, respectively. 
The Coulomb interaction strength is given by $U$. 
In Eq.~\eqref{eq:Hamiltonian} we made use of the operator identity $\hat n_\uparrow \hat n_\downarrow = \frac{1}{2} (\hat n_\uparrow + \hat n_\downarrow) - \frac{1}{2} (\hat n_\uparrow - \hat n_\downarrow)^2$ and incorporated the terms linear in $\hat n_\sigma$ by shifting the energy of the quantum dot's level, such that $\epsilon_{0,\sigma} = \epsilon_0 + \sigma B/2$ with $\epsilon_0 = E_0 + U/2$, where the bare energy level $E_0$ in the absence of magnetic field and Coulomb interaction can be tuned via a gate voltage. To write the interaction in terms of $\hat n_\uparrow - \hat n_\downarrow$ turns later out to be advantageous for the discrete Hubbard-Stratonovich transformation.
An electron with energy $\epsilon_{\alpha \vb{k}} = \epsilon_{\vb{k}} - \mu_\alpha$ in lead $\alpha = (\aL,\aR)$ and with momentum $\vb{k}$ is created or annihilated by the operators $\hat c^\dagger_{\alpha\vb{k}\sigma}$ and $\hat c_{\alpha\vb{k}\sigma}$, respectively. 
Finally, $t_\alpha$ denotes the tunnel coupling between lead $\alpha$ and the quantum dot. 
The tunnel coupling strength between quantum dot and lead $\alpha$ is given by $\Gamma_\alpha = 2\pi \abs{t_\alpha}^2 \rho(\epsilon^{\mathrm F}_{\alpha\vb{k}})$, where $\rho(\epsilon^{\mathrm F}_{\alpha\vb{k}})$ denotes the density of states of lead $\alpha$ at the Fermi level. 
We work in the wide-band limit, which usually is a good approximation in the stationary regime \cite{Covito_2018}. 
We also assume a symmetric setup, where $\Gamma = \Gamma_\aL = \Gamma_\aR$ and where the chemical potential of the leads are given by the bias voltage $\mu_\alpha = \pm eV/2$, for the left and right lead, respectively. 

\subsection{Path-integral formulation}

\begin{figure}[b]
\centering
\includegraphics[width=\columnwidth]{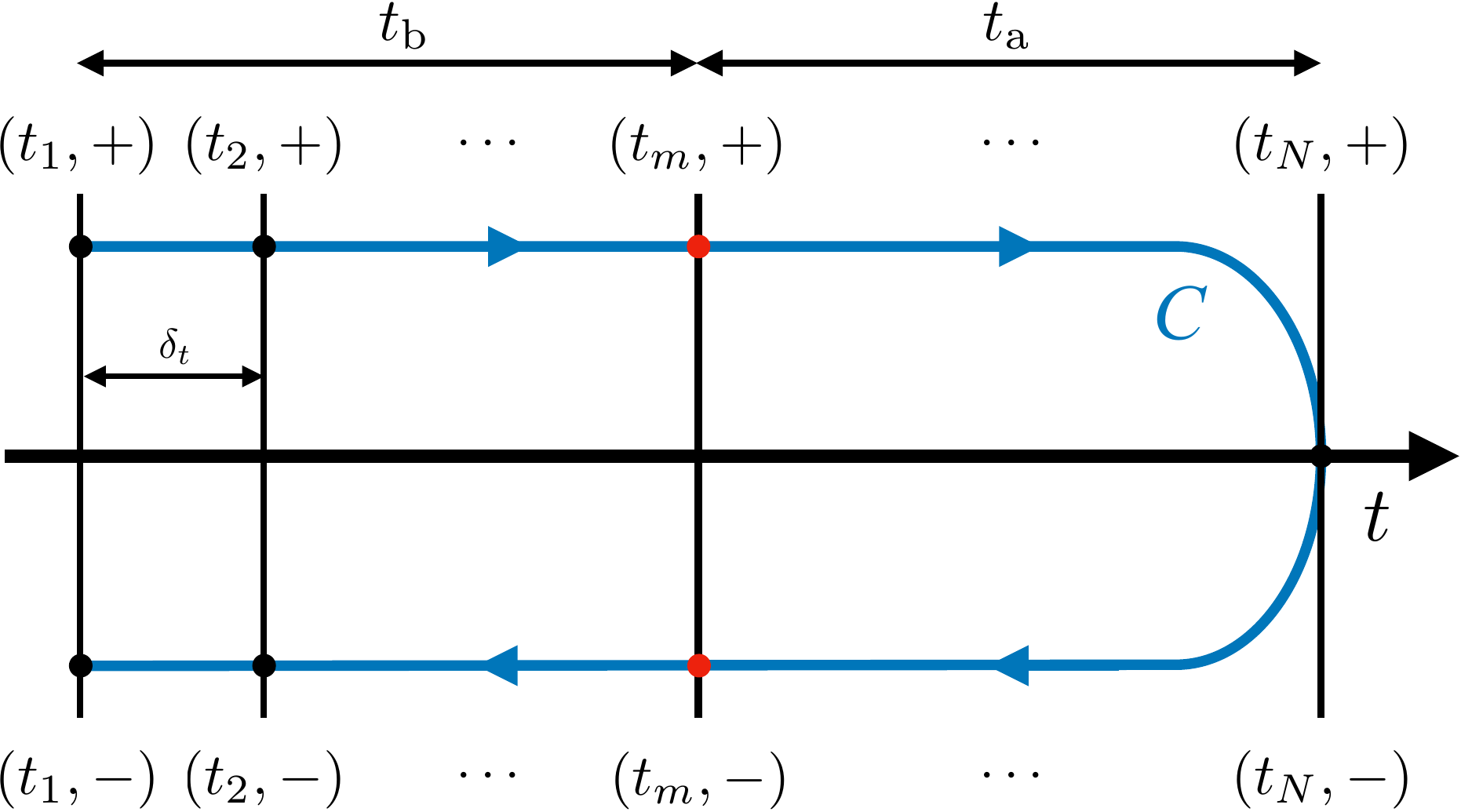}
\caption{The Keldysh contour $C$ with the measurement time $\tm$ at its center. 
The contour gets discretized into a total of $2N$ time steps of size $\delta_t$.
}
\label{fig:contour}
\end{figure}

We are interested in discussing time-local observables, namely the current, the quantum dot's occupation number and its spin-projector expectation value in the stationary limit.
The latter is achieved for $\tb \to \infty$, where $\tb=\tm-t_1$ (\textit{before}) is the time interval between the initialization and the measurement time $\tm$, while $\ta=t_N-\tm$ (\textit{after}) denotes the time interval between $\tm$ and the Keldysh return time $t_N$. 
Nonequilibrium properties are taken into account within a functional integral formulation on the Keldysh contour $C$ (see Fig.~\ref{fig:contour}) \cite{Kamenev, Negele_Orland, Mundinar_2019}. 
To take the forward and backward branch of the Keldysh contour into account, it is useful to define $\tau = (t,\nu)$, with physical time $t$ and Keldysh branch index $\nu=\pm$, with $+$ and $-$ representing the upper and lower Keldysh contour, respectively. 

For any time-local observable $\hat{O}$ at measurement time $\tm$, we introduce a source term $\eta O(\tm)$, with $\eta \in \mathds{R}$, in order to break time-translation symmetry on the level of the system's action $\mathcal{S}$ in a specific way. 
This allows us to calculate the expectation value of said observable via a derivative of the system's Keldysh generating functional,
\begin{align} \label{eq:GenExpVal}
	\expval*{\hat{O}} =  \pdv{\eta} \ln Z[\eta] \bigg|_{\eta=0}.
\end{align}
The Keldysh generating functional for the Hamiltonian given in Eq.~\eqref{eq:Hamiltonian} takes the form 
\begin{align}\label{eq:GenFuncDef}
	Z[\eta] = \int \mathcal{D}[d, c] \, \ee^{\ii \mathcal S + \eta O(\tm)},
\end{align}
and fulfills $Z[0]=1$ by construction.
The functional integral is built in the basis of fermionic coherent states $\ket{\Psi(\tau)}$, being defined via the eigenvalue equations of the fermionic annihilation operators
\bS\begin{align}
	\hat d_\sigma \ket{\Psi(\tau)} &= d_{\sigma} (\tau) \ket{\Psi(\tau)}, \\
	\hat c_{\alpha \vb{k} \sigma} \ket{\Psi(\tau)} &= c_{\alpha \vb{k} \sigma} (\tau) \ket{\Psi(\tau)}.
\end{align}\eS
As a result, the action $\mathcal S$ and the source term $\eta O(\tm)$ in Eq.~\eqref{eq:GenFuncDef} are functions of the Grassmann fields $\bar{d}_{\sigma} (\tau),d_{\sigma} (\tau), \bar{c}_{\alpha \vb{k} \sigma} (\tau), c_{\alpha \vb{k} \sigma} (\tau)$, and the functional integral runs over all of these degrees of freedom \cite{Kamenev, Mundinar_2019, Weiss_2013}. 
Dropping the explicit $\tau$-dependency, the action takes the form 
\begin{align}
    \mathcal{S} = & \int_C \dd{t} \left[ \sum_\sigma \bar{d}_\sigma (\ii \partial_t - \epsilon_{0,\sigma}) d_\sigma - \frac{U}{2} \left( n_\uparrow - n_\downarrow\right)^2 \right. \\
    & \left. + \sum_{\alpha \vb{k} \sigma} \bar{c}_{\alpha \vb{k} \sigma} ( \ii \partial_t - \epsilon_{\alpha \vb{k}}) c_{\alpha \vb{k} \sigma} 
    + \sum_{\alpha \vb{k} \sigma} \big( t_\alpha \bar{c}_{\alpha \vb{k} \sigma} d_\sigma + \text{h.c.}\big)\right]\nonumber
\end{align}

Performing the integral in Eq.~\eqref{eq:GenFuncDef} does not pose any challenge on the terms contributed by the leads' and the tunneling Hamiltonian, since they are quadratic in the Grassmann fields. 
However, the quartic on-site interaction term has to be tackled via a discrete Hubbard-Stratonovich (HS) transformation \cite{Hubbard_1959, Stratonovich_1958, Hirsch_1983}. 
For this, we first discretize the Keldysh contour into $2N$ time slices of length $\delta_t$ (see Fig.~\ref{fig:contour}), and then perform a discrete HS transformation,
\begin{align}\label{eq:HStrafo}
	 \ee^{-\frac{1}{2} \ii \nu \delta_t U \left( n_\uparrow-n_\downarrow \right)^2} 
	 = \frac{1}{2} \sum_{s=\pm 1} \ee^{-s\zeta_\nu (n_\uparrow-n_\downarrow)},
\end{align}
on each of these $2N$ slices, keeping in mind that $n_\sigma$ can only assume the values $0$ and $1$.
This decouples the interaction term, at the cost of introducing one Ising-like degree of freedom, $s=\pm 1$, per time slice. 
The HS parameter $\zeta_\nu$ is determined uniquely for $0\leq \delta_t U < \pi$ via \cite{Mundinar_2019, Weiss_2013}
\begin{align}
 	\cosh \zeta_\nu = \ee^{-\frac 1 2 \ii \nu \delta_t U}. 
\end{align}
As a result, we have introduced $2N$ new HS spins but are able to also integrate over the dot degrees of freedom, solving the functional integral in Eq.~\eqref{eq:GenFuncDef}. 
We find 
\begin{align} \label{eq:modGenFunc}
    \Zp[\eta] = \sum_{\vs} \det D[\eta,\vs\,],
\end{align}
with the discretized generating functional $\Zp[\eta]\propto Z[\eta]$, and with the matrix \cite{Weiss_2013,Mundinar_2019}%
\bS\label{eq:modDressedGF}
\begin{align} \label{eq:Ddef}
    D[\eta,\vs\,] & = S \left[ \Delta^{-1} - S \Sigma^C + \eta \Sigma^O\right] \Delta \\
    & = S - \Sigma^C \Delta +\eta S \Sigma^O \Delta \\ 
    & = S - \DC + \eta S \DO.
\end{align} 
\eS
For this we used the HS spin vector
\begin{align}\label{eq:vs}
    \vs = (s^+_1, s^-_1, s^+_2, s^-_2,\ldots, s^+_N, s^-_N),    
\end{align}
such that the sum includes all $2^{2N}$ possible spin configurations along the discretized Keldysh contour. 
In addition, we introduced several new matrices: 
We identify the inverse time-discrete Green's function $\Delta^{-1} = \Delta_0^{-1} - \Sigma^{\text{T}}$ of the non-interacting setup, where $\Delta_0$ is the free dot's Green function, and $\Sigma^\text{T}=\sum_\alpha \Sigma^{\text{T},\alpha}$ is the tunneling self energy. 
In addition, we introduced the charging self energy $S \Sigma^C$ with the diagonal spin matrix $S = \diag(\vs\,) \otimes \sigma_z$, which is the only part depending on the HS spins $\vs$, as well as the source self energy $\Sigma^O$, which is included to account for the source term. 
We also made use of the short-hands $\DC = \Sigma^C \Delta$ and $\DO = \Sigma^O \Delta$ for later convenience.
All of these matrices have dimensions $4N\times 4N$ due to the Trotter slicing (see Fig.~\ref{fig:contour}) and the spin degree of freedom. 

Note that we modified the generating functional by absorbing the factor $\frac{1}{2}$ per spin from Eq.~\eqref{eq:HStrafo}. 
Additionally, we have multiplied $S$ from the left and $\Delta$ from the right in Eq.~\eqref{eq:modDressedGF}.   
These changes do not affect the expectation value of the observable, since $\det S=1$ and $\det \Delta=\mathit{const}$, which cancels due to the logarithmic derivative in Eq.~\eqref{eq:GenExpVal}. 
However, multiplying by $\Delta$ ensures that $D[\eta,\vs\,]$ decays exponentially, while through the multiplication with $S$ the HS spins are located only on the diagonal of $D[\eta,\vs\,]$ as well as on the parts affected by the source self energy. 
The first property will be crucial when implementing the ISPI scheme, while the second is useful for an efficient implementation of the transfer-matrix formulation. 

\subsection{Form of the matrices}

To specify the elements of the matrices in Eq.~\eqref{eq:Ddef} we  employ the basis $(n,\nu,\sigma)$, where the Trotter index $n = 1,\ldots,N$ labels the time slice, the Keldysh index $\nu = \pm$ distinguishes the upper from the lower Keldysh contour, and $\sigma = \,\uparrow,\downarrow$ denotes the spin. 
In this basis, the Green's function of the non-interacting quantum dot in the presence of leads is given as \footnote{We use the notation $A=[a_{nn'}]_{nn'}$ to build the matrix $A$ from its elements $a$.} 
\begin{align}\label{eq:gom}
\Delta &= \bigg[ \int \frac{\dd{\omega}}{2\pi} \ee^{-\ii\omega (n-n')\delta_t} \times \\
&\quad\times\Big\{\sigma_z \otimes \left[\left(\omega - \epsilon_0\right)\sigma_0 - \tfrac{B}{2} \sigma_z\right] - \gamma_+(\omega)\otimes \sigma_0 \Big\}^{-1} \bigg]_{n n'},\nonumber
\end{align}
with $\gamma_\pm(\omega)=\gamma_\aL(\omega)\pm\gamma_\aR(\omega)$, where $\gamma_\alpha(\omega)$ denotes the $2\times 2$ Keldysh matrix
\begin{align}
\gamma_\alpha(\omega) =\frac{\ii}{2} \Gamma_\alpha  
\begin{pmatrix}
 2f_\alpha(\omega)-1 & -2f_\alpha(\omega) \\
-2f_\alpha(\omega)+2 &  2f_\alpha(\omega)-1
\end{pmatrix},
\end{align}
and $\sigma_z$ and $\sigma_0$ are the Pauli matrices, acting on either Keldysh or spin space if they appear on the left or on the right of the tensor product, respectively.
The Fermi function $f_\alpha(\omega)=[\exp(\beta(\omega-\mu_\alpha))+1]^{-1}$ describes the equilibrium occupation distribution of lead $\alpha$. 
Note that due to the symmetric discretization of the derivative $\partial_t \mapsto \omega^{-1}$ in frequency space, the discretized advanced and retarded Green's functions have the diagonal $|\Delta^{\mathrm{a,r}}|_{nn}=\frac{1}{2}$ instead of $1$, such that $\det(2\Delta)=1$.

The charging self energy $S \Sigma^C$ is time-local and therefore a diagonal matrix, with
\begin{align}\label{eq:ChargingSE}
\Sigma^C & = \diag\left[\ii \begin{pmatrix} \zeta_+ & 0 \\ 0 & \zeta_- \end{pmatrix} \otimes \sigma_0 \right]_{n}.
\end{align}

The form of the source self energy $\Sigma^O$ is based on the observable of interest $\hat{O}$, since it is derived from its source term $\eta O(\tm)$. 
The source self energies for the occupation number, the spin projection in $z$ direction, and for the current were derived in earlier works \cite{Mundinar_2020, Weiss_2008, Weiss_2013, Mundinar_2019}. 
As a result, we only present the results here, and refer to the aforementioned references for a more detailed derivation. 
Assuming we include the measurement on the Trotter slice $m=\tm/\delta_t$, we find for the occupation number $\hat N = \sum_\sigma \hat n_\sigma$ and for the spin projection $\hat S_z = \frac{1}{2} (\hat n_\uparrow - \hat n_\downarrow)$ \cite{Mundinar_2020}
\bS\label{eq:SourceSE} 
\begin{align}\label{eq:SourceLoc}
	\Sigma^{N} &= \left[\,\delta_{n m} \delta_{n'm} \begin{pmatrix} 0 & 0 \\ 1 & 0 \end{pmatrix} \otimes \sigma_0\,\right]_{n n'}, \\
	\Sigma^{S_z} &= \left[\,\delta_{n m} \delta_{n' m} \begin{pmatrix} 0 & 0 \\ 1 & 0 \end{pmatrix} \otimes \frac{\sigma_z}{2}\,\right]_{n n'}. 
\end{align}
These are sparse matrices, containing only two nonzero elements. 
Finally, the current operator is given by $\hat I = -\ii e/2 \sum_{\alpha \vb{k} \sigma} \alpha t_\alpha (\hat c^\dagger_{\alpha \vb{k} \sigma} \hat d_\sigma - \hat d^\dagger_\sigma \hat c_{\alpha \vb{k} \sigma})$, with its source term given by \cite{Weiss_2008, Weiss_2013, Mundinar_2019} 
\begin{align}\label{eq:SourceI}
    \Sigma^{I} = \frac{e}{2} \Re\! \left[\ii \delta_{n m}  \int \frac{\dd \omega}{2\pi} \frac{\left[\sigma_z\gamma_-(\omega)\right] \otimes \sigma_0}{\ee^{\ii \omega(n-n') \delta_t}}\right]_{n n'}.
\end{align}\eS
Note that the source self energy for the current operator is a sparse matrix, too, with only one row $m$ having non-vanishing elements. 

\section{Method} \label{sec:Method}

\subsection{Truncation of matrix \textit D}
\begin{figure}[t]
\centering
\includegraphics[width=\columnwidth]{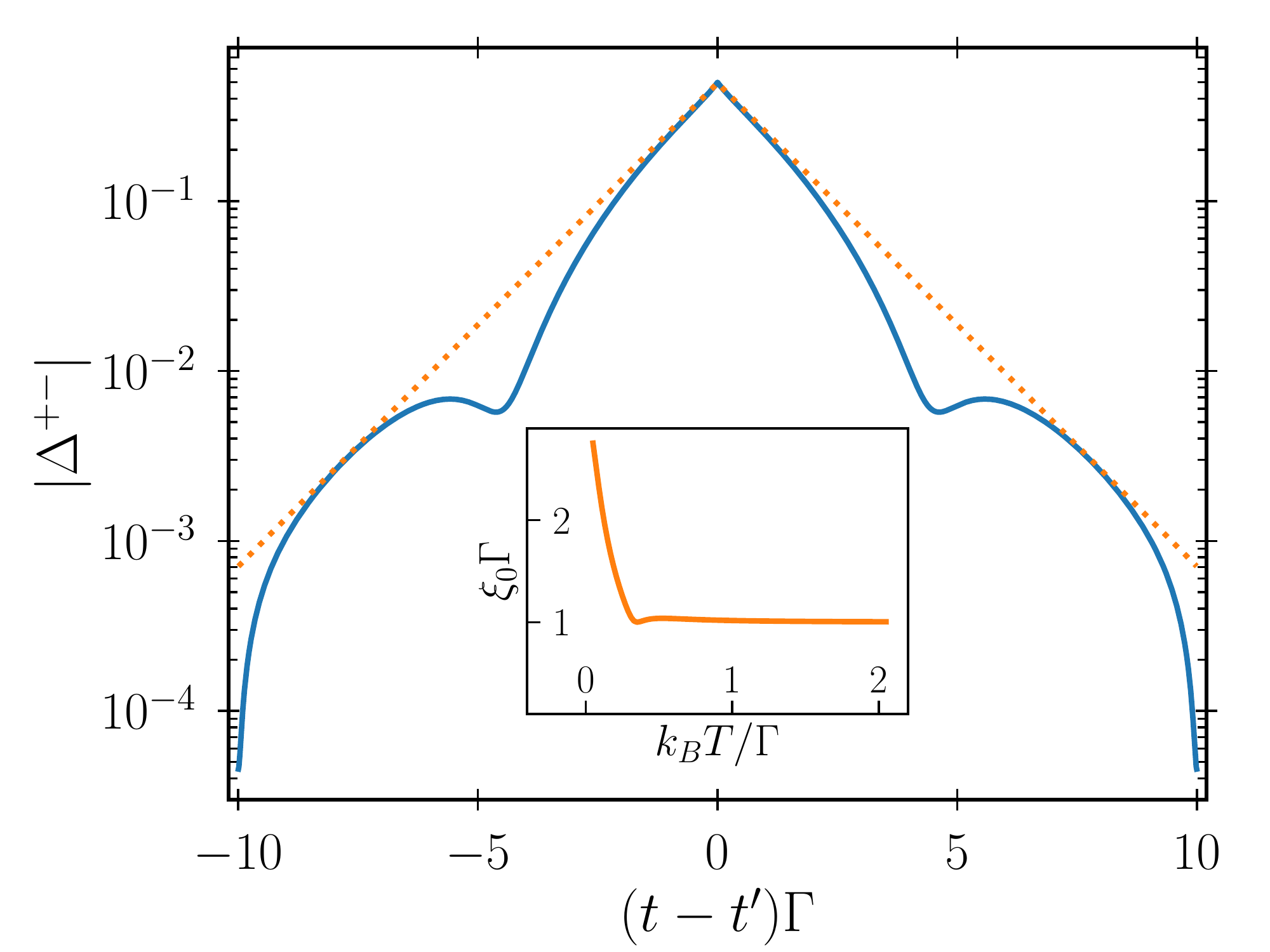}
\caption{The system's dressed, non-interacting Green's function $\Delta$ as a function of time $(t-t')$ for parameters $\kB T=0.2\,\Gamma$, $\epsilon_0=0$, $eV=\Gamma$, $B=0$. 
The dotted line is an exponential fit for the envelope function according to $|\Delta|\propto \exp(-|t-t'|/\xi_0)$.
Inset: Correlation time $\xi_0$ of the envelope function for the non-interacting system as a function of temperature $\kB T$. 
Other parameters are the same as in the main panel.}
\label{fig:GFdecay}
\end{figure}

The sum in Eq.~\eqref{eq:modGenFunc} runs over all configurations of the $2N$ HS spins, with $N$ usually being of the order of several hundreds. 
Summing over these $2^{2N}$ configurations is an insurmountable task, and approximations are in order. 
The approximation for the ISPI scheme is based on the fact that lead-induced correlations decay exponentially with time at finite temperatures \cite{Weiss_2008, Weiss_2013}. 
As a consequence the system's non-interacting Green's function $\Delta$ -- and with it $D[\eta,\vs\,]$ -- also decays exponentially with time $|t-t'|$. 
This is shown in the main panel of Fig.~\ref{fig:GFdecay}, where the absolute value of $\Delta^{+-}$ is plotted against $|t-t'|$ for the parameter set $\kB T = 0.2\,\Gamma$, $\epsilon_0=0$, $eV=\Gamma$, $B=0$ and $\delta_t \to 0$. 
While for low temperatures one finds large oscillations as seen in the figure, the enveloping function still decays exponentially, as long as $\kB T >0$. 
We demonstrate this temperature dependence of the correlations' decay in the inset of Fig.~\ref{fig:GFdecay}, where the correlation time $\xi_0$ of the enveloping function is plotted as a function of temperature.

This motivates us to truncate the interacting Green's function Eq.~\eqref{eq:modDressedGF}, such that 
\begin{align}\label{eq:tridiagD}
D[\eta,\vs\,] = 
\begin{bmatrix}
 D_{1}(\vs_1) & D_{1}^+      &         & \\
 D_{2}^-      & D_{2}(\vs_2) & D_{2}^+ & \phantom{\ddots} \\
              & D_{3}^-      & \ddots  & \ddots  \\
              &              & \ddots  & D_{L}(\vs_{L})
\end{bmatrix}
\end{align}
is a block-tridiagonal matrix with $4K \times 4K$ blocks $D_{\ell}^{(\pm)}$, where $\ell=1,\ldots,L$ and $L = N/K$. 
The number of Trotter slices within one block, $K = t_K/\delta_t$, depends on a chosen memory time $t_K$ and the Trotter step size $\delta_t$.
Since only row $m$ in Eq.~\eqref{eq:modDressedGF} is affected by the source self energy, see Eqs.~\eqref{eq:SourceSE}, this also holds true for the block matrix, where only the block row $\mm$ containing row $m$ is affected by the source self energy. 
We also introduced the block spins $\vs_\ell$, each consisting of a set of $2K$ HS spins, such that $\vs=(\vs_1,\ldots,\vs_L)$, see Eq.~\eqref{eq:vs}.

Using this truncation we are now able to formulate the ISPI scheme, which was first applied to the Anderson model in Ref.~\onlinecite{Weiss_2008} and later employed to study the Anderson-Holstein model \cite{Huetzen_2012}, and the quantum-dot spin valve \cite{Mundinar_2019, Mundinar_2020}. 
We introduce the main ideas in the next section, and refer to any of the aforementioned references for a more complete discussion. 
After that we introduce the mapping of the ISPI method to a transfer-matrix method, which by construction directly addresses the stationary limit and drastically increases the computational performance.

\subsection{ISPI formulation} \label{sec:FTImplementation}
As can be seen in Eq.~\eqref{eq:modGenFunc} one has to calculate the determinant of the block-tridiagonal matrix from Eq.~\eqref{eq:tridiagD}. 
According to Ref.~\onlinecite{Salkuyeh_2006} such a determinant can be evaluated iteratively via 
\begin{align}\label{eq:GenFuncDcheck}
 	\Zp_{L}[\eta] = \sum_{\vs} \prod_{\ell=1}^{L} \det \check{C}_{\ell}(\vs_{1:\ell}),
\end{align}
where $\vs_{1:\ell}=\{\vs_{1},\ldots,\vs_\ell\}$, and
\begin{align}\label{eq:Dcheck}
 	\check{C}_{\ell}(\vs_{1:\ell}) &= D_{\ell}(\vs_\ell) - D_{\ell}^- \, \check{C}_{\ell-1}^{-1}(\vs_{1:\ell-1}) \, D_{\ell-1}^+
\end{align}
denotes the Schur complement in the $\ell$-th step, with $\check{C}_{1}(\vs_{1:1}) = D_{1}(\vs_1)$ and $\ell = 2,\ldots, L$. 
Therefore, each $\check{C}_{\ell}$ depends on all previous $\check{C}_{k}$, with $k<\ell$, and since the spins are distributed diagonally, Eq.~\eqref{eq:Dcheck} connects $\vs_\ell$ with all previous $\vs_k$. 
To remain consistent with the idea of truncating correlations after the memory time $t_K$, we approximate $\check{C}_{\ell}(\vs_{1:\ell})$ by \cite{Weiss_2008, Mundinar_2019} 
\begin{align}\label{eq:Dtilde}
 	C_{\ell}(\vs_{\ell-1},\vs_{\ell}) &= D_{\ell}(\vs_\ell) - D_{\ell}^- \, D_{\ell-1}^{-1}(\vs_{\ell-1}) \, D_{\ell-1}^+,
\end{align}
that is, in Eq.~\eqref{eq:Dcheck} we replace $\check{C}_{\ell-1}^{-1}(\vs_{1:\ell-1})$ with $D_{\ell-1}^{-1}(\vs_{\ell-1})$, which only depends on $\vs_{\ell-1}$. 
Therefore,
$C_{\ell}$ only connects $\vs_{\ell-1}$ and $\vs_{\ell}$, effectively truncating interaction-induced correlations after the memory time $t_K$. 
As a result, we are able to rewrite Eq.~\eqref{eq:GenFuncDcheck}, finding
\begin{align}\label{eq:GenFuncDtilde}
 	\Zp_{L}[\eta] &= \sum_{\vs_{1}} \det D_{1}(\vs_1) \sum_{\vs_{2}} \det C_{2}(\vs_{1},\vs_{2}) \times \cdots \nonumber \\
 	& \quad \times \sum_{\vs_{L}} \det C_{L}(\vs_{L-1},\vs_{L}).
\end{align}
Note that we used the fact that each $C_{\ell}$ depends only on the two block spins $\vs_{\ell-1}$ and $\vs_{\ell}$, allowing us to evaluate $L$ sums over $M=2^{2K}$ spin configurations, instead of the much larger sum over $2^{2N}$ configurations. 
Since we can choose $K\ll N$, this is a huge reduction of complexity, allowing us to evaluate the generating functional $\Zp[\eta]$ for much larger values of $N$. 

ISPI can be characterized as a finite-time implementation due to the finite length of the considered Keldysh contour as depicted in Fig.~\ref{fig:contour}.
In order to reach the stationary limit, the time $\tb$ before the measurement has to be chosen large enough such that the system has relaxed from any arbitrary initial state to its stationary state. 
This can be ensured by choosing $\Gamma \tb \gg 1$, depending on the system under consideration. 
For the Anderson model, we find that a time interval of $\Gamma \tb = 15$ is sufficient. 

The finite-time implementation has the disadvantage that, first, one needs to carefully choose for each calculation the proper time interval to make sure that the stationary limit has been achieved.
Second, increasing the time interval (due to larger relaxation times) increases the computational cost.
The TraSPI formulation, put forward in this paper, is motivated by the desire to directly perform the limit $\tb\to\infty$ analytically by using transfer matrices.
This strategy does not only strongly decrease the computational cost but has also other benefits, as described in the next section.

\subsection{TraSPI formulation} \label{sec:TMImplementation}
The term ``transfer matrix''  is reminiscent of the transfer-matrix method known from statistical physics, which was first introduced to solve the 1-dimensional Ising model \cite{Kramers_1941, Wannier_1941} and later used by Onsager as the basis for his well-known exact solution of the two-dimensional Ising model \cite{Onsager_1944, Kaufman49}, for a recent application see, e.g., \cite{Hucht16a,Hucht21a}.
In fact, Eq.~\eqref{eq:GenFuncDtilde} is the basis for a mapping of the ISPI scheme to a transfer-matrix formulation, which we derive here.  

\begin{figure}[t!]
\centering
\includegraphics[width=0.9 \columnwidth]{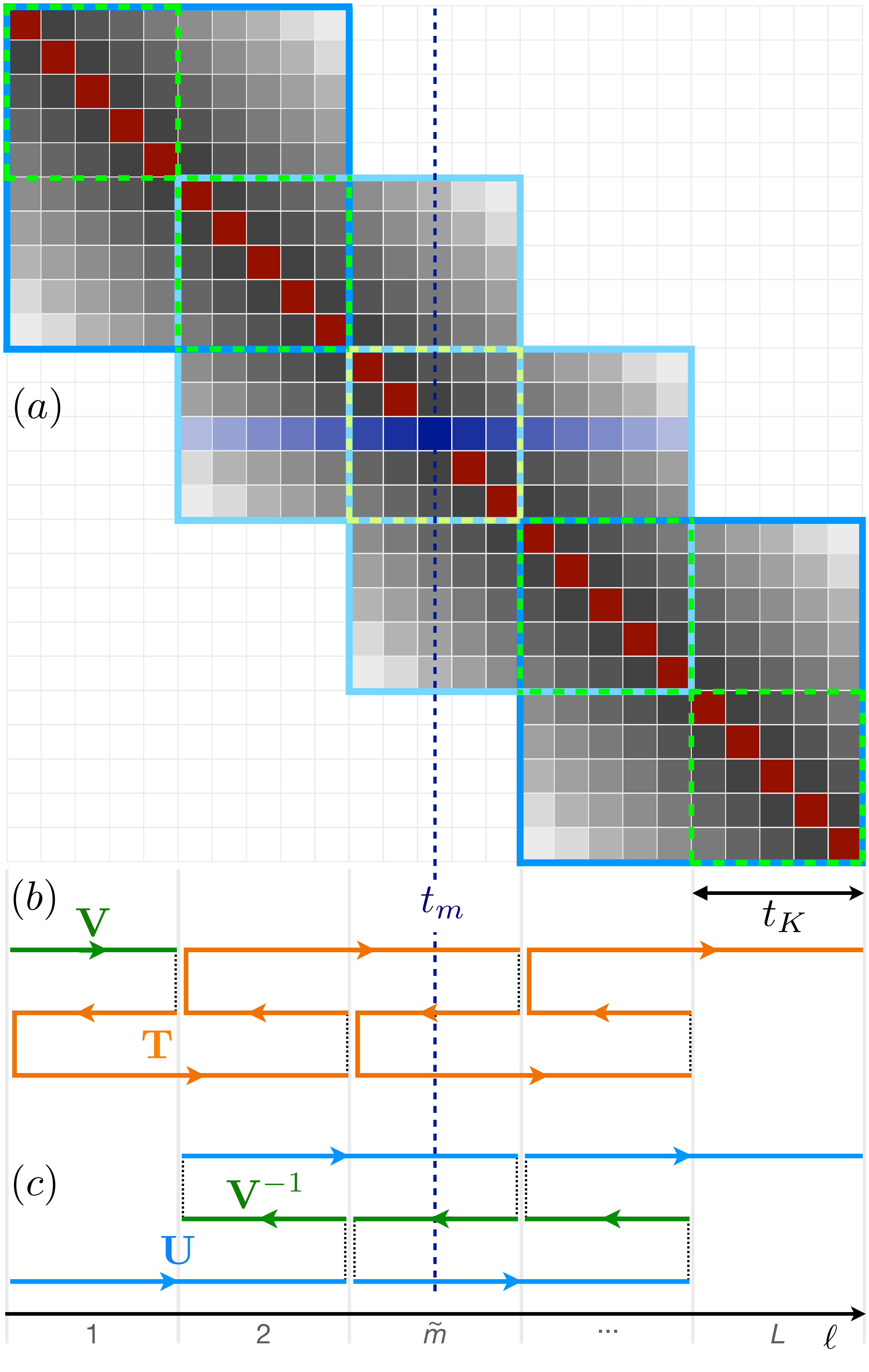}
\caption{(a) Sketch of the elements of $D[\eta,\vs\,]$ after the truncation Eq.~\eqref{eq:tridiagD}, for the case $L=K=5$. 
Small boxes represent $4\times 4$ matrices spanned by the Keldysh and spin space, while saturation depicts the absolute value of the elements. 
Dark red boxes carry HS spins $s^\nu_n$, blue boxes are affected by the source self energy and carry HS spins $s^\nu_{m}$, grey boxes are neither affected by the source term nor by the HS spins. 
The blue solid borders represent elements included in the TM $\TMU_{\ell-1,\ell}$, whereas
the green dashed borders represent elements included in the TM $\TMV_{\!\ell}$. 
For these, lighter colors denote TMs affected by the source term at $\tm$. 
(b) and (c) are graphical representations of Eqs.~\eqref{eq:AsymTransGenFunc} and \eqref{eq:SymTransGenFunc}, respectively. 
The TM $\TM$ (orange) first propagates one step of size $t_K$ backwards in time, followed by two steps in forward direction.
The TM $\TMU$ (blue) propagates two steps of size $t_K$ forward in time, while $\TMV$ and $\TMV^{-1}$ (green) are propagating one step forward or backward, respectively.}
\label{fig:TMImplement}
\end{figure}

First, we reiterate that each $C_{\ell}$ depends on two block spins $\vs_{\ell-1}$ and $\vs_{\ell}$, while $D_{1}$ depends only on $\vs_1$. 
Therefore, the shape of Eq.~\eqref{eq:GenFuncDtilde} suggests rewriting it into a matrix product in the space of HS-spin configurations. 
For this, we enumerate the $M$ different configurations of the block spins $\vs_\ell$ by $\mu=0,\ldots,M{-}1$, such that, e.g., $\mu=0$ corresponds to $2K$ HS spins pointing up, $\vs_\ell=(1,\ldots,1)$.
In addition, we write $f_\ell(\mu)$ instead of $f_\ell(\vs_\ell)$ for any function that depends on the HS spins, in order to keep the notation compact. 
With this, we define the $M\times M$ transfer matrix (TM)
\begin{align}\label{eq:LambdaDef}
 	\TM_{\ell-1,\ell} &= \big[\det C_{\ell}(\mu,\mu') \,\big]_{\mu\mu'}.
\end{align}
Thus, each row corresponds to one of the $M$ configurations $\mu$ of the block spin $\vs_{\ell-1}$ and each column to one of the $M$ configurations $\mu'$ of $\vs_{\ell}$. 
If we additionally define the two vectors
\bS\begin{align}
	\bra{v} &= \big[ \det D_{1}(\mu) \,\big]_\mu,\\
	\ket{1} &= \big[ 1 \big]_\mu,
\end{align}\eS
then the Keldysh generating functional \eqref{eq:GenFuncDtilde} takes the simple matrix-product form
\begin{align}\label{eq:AsymTransGenFunc}
	\Zp_{L}[\eta] = \mel*{v}{\TM_{1,2}\TM_{2,3}\cdots\TM_{L-1,L}}{1},
\end{align}
and each multiplication with $\TM_{\ell-1,\ell}$ corresponds to one term $\sum_{\vs_{\ell}} \det C_{\ell}(\vs_{\ell-1},\vs_{\ell})$
in Eq.~\eqref{eq:GenFuncDtilde}.

In the next step, we symmetrize Eq.~\eqref{eq:AsymTransGenFunc} by introducing two new transfer matrices $\TMU_{\ell-1,\ell}$ and $\TMV_{\!\ell}$. 
For this, we reinspect the definition \eqref{eq:Dtilde} of $C_{\ell}$, whose determinant provides the elements of $\TM_{\ell-1,\ell}$, noting that it has the well-known form of the Schur complement of a part of the tridiagonal block matrix $D[\eta,\vs\,]$, given by
\begin{align}\label{eq:Dbar}
 	D_{\ell-1,\ell}(\mu,\mu') = \begin{bmatrix}
 	D_{\ell-1}(\mu) & D_{\ell-1}^+ \\
 	D_{\ell}^-      & D_{\ell}(\mu') \end{bmatrix}.
\end{align}
This is a $2\times 2$ block matrix, affected by the block spins $\vs_{\ell-1}$ and $\vs_{\ell}$. 
Thus, we are able to rewrite $\det C_{\ell}$ as the quotient of two determinants,
\begin{align}\label{eq:DtildeSchur}
	\det C_{\ell}(\mu,\mu') = \frac{\det D_{\ell-1,\ell}(\mu,\mu')}
	{\det D_{\ell-1}(\mu)}.
\end{align}
Returning to the transfer-matrix formalism, this corresponds to the matrix product
\begin{align}
    \TM_{\ell-1,\ell} = \TMV^{-1}_{\!\ell-1} \TMU^{\vphantom{1}}_{\ell-1,\ell} \, , 
\end{align}
where $\TMU_{\ell-1,\ell}$ is a dense $M\times M$ transfer matrix and $\TMV_{\!\ell}$ is a $M\times M$ diagonal matrix,
\bS\label{eq:UVDef}
\begin{align}
    \TMU_{\ell-1,\ell} &= \big[ \det D_{\ell-1,\ell}(\mu,\mu') \,\big]_{\mu\mu'} \\
    \TMV_{\!\ell} &= \diag\!\big[ \det D_{\ell}(\mu) \,\big]_{\mu}.
\end{align}
\eS
Building the generating functional with these, we get a symmetrized version of Eq.~\eqref{eq:AsymTransGenFunc},
\begin{align}\label{eq:SymTransGenFunc}
    \Zp_{L}[\eta] = \mel**{1}{\TMU^{\vphantom{1}}_{1,2} \TMV^{-1}_{\!2} \TMU^{\vphantom{1}}_{2,3} \cdots  \TMV^{-1}_{\!L-1} \TMU^{\vphantom{1}}_{L-1,L}}{1},
\end{align}
where we used the fact that 
$\bra{v}= \bra{1} \TMV_{\!1}$.
We emphasize that to this point the Eqs.~\eqref{eq:GenFuncDtilde}, \eqref{eq:AsymTransGenFunc}, and \eqref{eq:SymTransGenFunc} are synonymous, and no additional assumptions have been made. 
In Fig.~\ref{fig:TMImplement} we show a sketch of the elements of $D[\eta,\vs\,]$ that are taken into account for the case $L=K=5$. 
Each of the small $1\times 1$ boxes represents a single time slice of $D[\eta,\vs\,]$ (thus a $4\times 4$ block, to take spin and Keldysh into account.).

The number of matrices entering Eq.~\eqref{eq:SymTransGenFunc} increases linearly with $L$.
To achieve the stationary limit, we have to choose $L$ large.
We now improve upon this: (i) we perform the stationary limit analytically, (ii) we optimize the position of the measurement, (iii) we calculate the derivative of $\pdv{\eta}\Zp_{L}[\eta]$ at $\eta=0$ analytically, and (iv) we implement a differential measurement method. 

\subsubsection{Stationary limit}
\begin{figure}[t]
\centering
\includegraphics[width=\columnwidth]{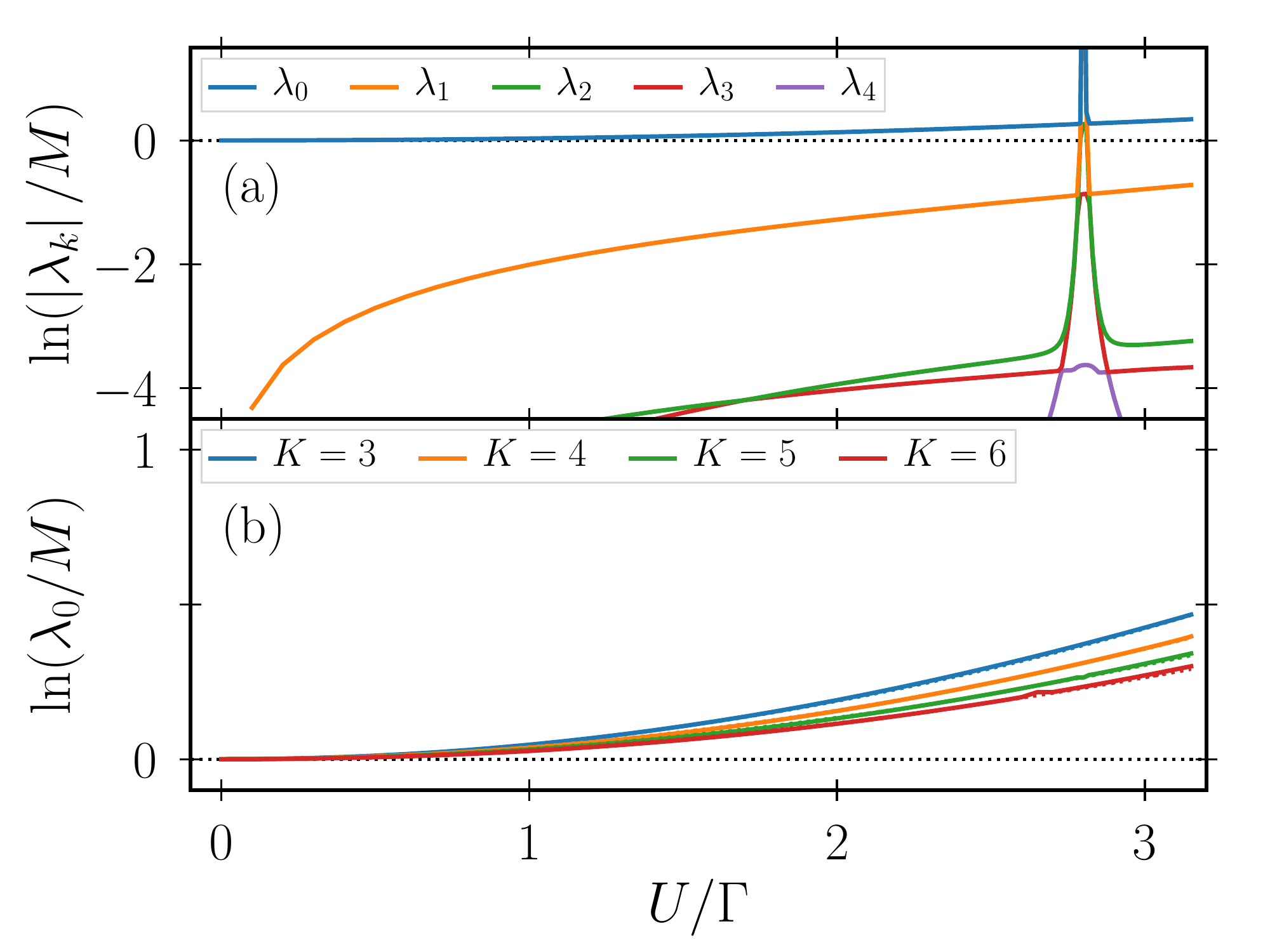}
\caption{(a) The absolute value of the five leading eigenvalues of the TM $\TM$ as a function of the interaction strength $U$ for $\Gamma t_K = 1.5$ and $K=5$. 
(b) The eigenvalue $\lambda_0$ as a function of $U$ for $\Gamma t_K = 1.5$ and for different $K$ (and therefore $\delta_t$). 
Dotted lines show the approximation \eqref{eq:lnlambda}.
Shown is the parameter set given by $\kB T=0.5\,\Gamma$, $\epsilon_0=0$, $eV=0.1\,\Gamma$, $B=0$.
Different parameters produce similar pictures.}
\label{fig:EigenvalueDevelopment}
\end{figure}
From this point on we are interested in the case that the system reached the stationary limit, i.e., we perform the limit $\tb\to \infty$. 
In order to perform this limit, we start with an index shift by $\mm$ in the block indices $(\ell,\mm)_\text{old}\mapsto (\ell,\mm)_\text{new}=(\ell-\mm,0)$, such that the measurement block becomes $\mm=0$ and $\ell=-\Lb-1,\ldots,\La+1$, with $\Lb+\La=L-3$. 
Note that $\Lb$ and $\La$ denote the number of TMs $\TM$ \textit{before} and \textit{after} the measurement, respectively.

In the next step, we make use of the fact that the system is symmetric under time-translation as long as the source term is not included, i.e., as long as $\eta = 0$. 
As a consequence, the non-interacting Green's function $\Delta$ obeys $\Delta_{n,n'} = \Delta_{n+1,n'+1}$. 
Thus, for the tridiagonal matrix $D[0, \vs\,]$ we equally find that $D_\ell(\mu)$ and $D_\ell^\pm$ are independent of $\ell$. 
With this and Eqs.~\eqref{eq:Dtilde}, \eqref{eq:Dbar}, we find that due to time-translation symmetry the transfer matrices fulfill $\TM_{\ell-1,\ell} = \TM$, $\TMU_{\ell-1,\ell} = \TMU$, and $\TMV_{\!\ell} = \TMV$ for all $\ell\neq 0, 1$. 

However, the source term was implemented to specifically break this time-translation symmetry, and therefore whenever an element of $D[\eta,\vs\,]$ is affected by the source self energy $\Sigma^O$, above relations do not hold anymore. 
We mentioned before that for time-local observables, the source self energy affects only the matrices $D_{0}^{\pm}$ and $D_{0}(\vs_{0})$ at most.
Consequently, only the transfer matrices acting at position $\mm=0$, i.\,e., $\TM_{-1,0}$, $\TM_{0,1}$, $\TMU_{-1,0}$, $\TMU_{0,1}$ and $\TMV_{\!0}$, are affected by the source term and are therefore different from the others, see Fig.~\ref{fig:TMImplement}. 
We find
\begin{align} \label{eq:LambdaGenFunc}
	\Zp_{L}[\eta] &= \mel*{v}{\TM^{\Lb} \TM_{-1,0} \TM_{0,1} \TM^{\La}}{1}.
\end{align}

To reach the stationary limit $\tb \to \infty$, we now can take the limit $L\to \infty$, meaning we let both $\Lb\to \infty$ and $\La\to \infty$.
For high powers of $\TM$ we use the identity
\begin{align}\label{eq:MatAsEigvec}
	\lim_{n\to\infty} \frac{\TM^n}{\lambda_0^n} 
	= \lim_{n\to\infty} \sum_{k\geq 0} \frac{\lambda_k^n}{\lambda_0^n} \dyad{\lambda_k} 
	= \dyad{\lambda_0},
\end{align}
where $\lambda_k$ are the eigenvalues of $\TM=\TMV^{-1}\TMU$, with $\abs{\lambda_0} > \abs{\lambda_1} \geq \ldots$, while $\bra{\lambda_k}$ and $\ket{\lambda_k}$ are the respective left and right eigenvectors, fulfilling $\braket{\lambda_k}{\lambda_{k'}}=\delta_{kk'}$. 
Likewise, we define $\bra*{\tilde\lambda_k}=\bra*{\lambda_k}\TMV^{-1}$ and
$\ket*{\tilde\lambda_k}=\TMV\ket*{\lambda_k}$ as corresponding left and right eigenvectors of $\TMU \TMV^{-1}$ (with the same eigenvalues, $\tilde\lambda_k=\lambda_k$).
Plugging this back into Eq.~\eqref{eq:LambdaGenFunc} and performing the stationary limit, we find
\bS\label{eq:LambdaGenFunc2}
\begin{align}
	\frac{Z_{\infty}[\eta]}{Z_{\infty}[0]} & = \lambda_0^{-2} \mel**{\lambda_0}{\TM_{-1,0} \TM_{0,1}}{\lambda_0} \\
\label{eq:UVGenFunc2}
	&= \lambda_0^{-2} \mel*{\tilde\lambda_0}{\TMU_{-1,0} \TMV^{-1}_{\!0} \TMU_{0,1}}{\lambda_0}.
\end{align}
\eS

In Fig.~\ref{fig:EigenvalueDevelopment}(a) we plot the absolute value of the largest five eigenvalues of the TM $\TM$ as a function of the Coulomb interaction strength $U$ for the parameter set $\kB T=0.5\,\Gamma$, $\epsilon_0=0$, $eV=0.1\,\Gamma$, $B=0$,  $\Gamma t_K = 1.5$, and $K=5$. 
We find that the largest eigenvalue is $\lambda_0=M$ at $U=0$, as would be expected, since in the noninteracting limit the transfer matrix $\TM$ is just a $M \times M$ matrix, where each element is equal to 1. 
For finite Coulomb interaction it scales with $\dtU = \delta_t U$ as
\begin{align}\label{eq:lnlambda}
    \ln\frac{\lambda_0}{M} = \frac{3(K-1)}{32}  \dtU^2 + \mathcal{O}(\dtU^3)
\end{align} 
for small $\dtU$. 
However, once $U$ becomes large, we find occasionally peaks where one of the lower eigenvalues diverges. 
These divergences are not physical but are a consequence of the discretization and truncation of the Green's function. 
We can track them down to elements of the diagonal matrix TM $\TMV$ that vanish, resulting in divergences in $\TMV^{-1}$, and hence in $\TM$. 
Nonetheless, the physical correct eigenvalue $\lambda_0$ is still present, and we choose that one (instead of the peaks) to calculate the correct generating functional. 
To choose the correct eigenvalue $\lambda_0$ we make use of the fact that the corresponding right eigenvector is $\ket{\lambda_0} = \ket{1} + \mathcal{O}(\dtU^3)$ (see below). 
Thus, we identify the correct eigenvalue by maximizing the overlap of the corresponding right eigenvector with $\ket{1}$. 
In Fig.~\ref{fig:EigenvalueDevelopment}(b) we plot this physically correct eigenvalue $\lambda_0$ as a function of $U$ for different $K$. 
As can be seen, only for $U \gtrsim 3\,\Gamma$ the data gets noisy, but increasing $K$ allows for the calculation of larger $U$. 

The TraSPI formulation is a significant reduction in complexity compared to the traditional, i.e., finite-time ISPI implementation: Instead of $L-1$ dense transfer matrices, we only have to evaluate the three dense TMs $\TMU$, $\TMU_{-1,0}$ and $\TMU_{0,1}$, as well as the two diagonal TMs $\TMV$ and $\TMV_{\!0}$. 
An additional benefit of the $\TMU\TMV$-decomposition \eqref{eq:UVDef} is that it allows for an analytic evaluation of the derivative with respect to $\eta$. 
Before, however, we demonstrate that we are able to shift the measurement time to the end of the Keldysh contour in order to reduce the number of necessary transfer matrices even further.

\subsubsection{Position of measurement}

In Eqs.~\eqref{eq:LambdaGenFunc2}, we assumed that the measurement is placed somewhere in the middle of the Keldysh contour (see also Fig.~\ref{fig:contour}). 
However, based on causality, whatever happens at physical times after the measurement must not have an impact on the outcome of the measurement itself. 
In other words, the time propagation $\ta$ along the Keldysh contour from $\tm$ to $t_N$ and back to $\tm$ is unitary and should therefore cancel out, see Fig.~\ref{fig:contour}.
As a consequence, it should not be detrimental to shift the measurement time forward in time on the Keldysh contour, until it is located at the rightmost point. 
In the words of the TM formulation from Eq.~\eqref{eq:LambdaGenFunc}, it should be sufficient to let $\Lb\to\infty$ and set $\La=0$, or when taking a look at the results from Eqs.~\eqref{eq:LambdaGenFunc2}, the vector $\ket{1}$ should be a right eigenvector of the TM $\TM$.

However, this exact unitarity present in the continuum limit \eqref{eq:GenFuncDef} is violated by the Trotter discretization.
Nevertheless, the error in the right eigenvector is quite small, $\ket{\lambda_0}=\ket{1}+\mathcal{O}(\dtU^3)$, and can be safely neglected, while the other eigenvectors show a stronger dependency on $\dtU$.
This means that we are able to rewrite Eqs.~\eqref{eq:LambdaGenFunc2} as
\bS
\begin{align}\label{eq:LambdaGenFuncEnd}
	\frac{Z_{\infty}[\eta]}{Z_{\infty}[0]} & = \lambda_0^{-2} \mel**{\lambda_0}{\TM_{-1,0} \TM_{0,1}}{1} \\
	&= \lambda_0^{-2} \mel*{\tilde\lambda_0}{\TMU_{-1,0} \TMV^{-1}_{\!0} \TMU_{0,1}}{1}.
\end{align}
\eS
Note that with this equation, the measurement takes place in the second to last $4K\times 4K$ block. 
If we actually measure on the last possible Trotter slice, $\tm = t_N$, only a single TM remains that is affected by the source term, resulting in the even simpler expression
\bS\label{eq:LambdaGenFuncEnd2}
\begin{align}
	\frac{Z_{\infty}[\eta]}{Z_{\infty}[0]} & = \lambda_0^{-1} \mel**{\lambda_0}{\TM_{-1,0}}{1} \\
\label{eq:UVGenFuncEnd2}
	& = \lambda_0^{-1} \mel*{\tilde\lambda_0}{\TMU_{-1,0}}{1},
\end{align}
\eS

This means, we are able to further reduce the number of necessary TMs to the two fully occupied TMs $\TMU$ and $\TMU_{-1,0}$, and one diagonal TM $\TMV$. 
We now turn to deriving the analytic derivative with respect to $\eta$, necessary to calculate observables from Eqs.~\eqref{eq:LambdaGenFuncEnd2}.

\subsubsection{Analytic derivative}

To calculate expectation values of observables via Eq.~\eqref{eq:UVGenFuncEnd2}, we make again use of Eq.~\eqref{eq:GenExpVal}. 
We explicitly calculate the derivative with respect to $\eta$ here. 
If for some reason we do not wish to position the measurement at the end of the Keldysh contour, meaning we use Eq.~\eqref{eq:UVGenFunc2} instead of Eq.~\eqref{eq:UVGenFuncEnd2}, calculations for the derivatives of the two remaining TMs work analogous as presented here. 
Performing the derivative yields
\begin{align}\label{eq:AnaDerivU}
    \expval*{\hat{O}} & = \frac{Z_\infty'[0]}{Z_\infty[0]} = \lambda_0^{-1} \mel*{\tilde\lambda_0}{\TMU_{-1,0}'}{\lambda_0},
\end{align}
where we wrote $A'$ for $\pdv{\eta}A|_{\eta=0}$.
The derivative of a matrix acts on its elements, which in this case are themselves determinants of $D_{\ell-1,\ell}$. 
Therefore, we make use of the identity $(\ln \det A)' = (\det A)'/\det A = {\tr}(A^{-1} A')$, and extend the notation introduced for $D_{\ell-1,\ell}$, Eq.~\eqref{eq:Dbar}, to other matrices, meaning $A_{\ell-1,\ell}$ denotes a $2\times 2$ part of a $L\times L$ block matrix $A$. 
We further simplify the notation by writing $A_{[2]}=A_{\ell-1,\ell}$ if $\ell \neq 0,1$, which are $2\times 2$ block matrices not affected by the source term. 
With this we find for the derivative \footnote{Note that we can safely write $\tr(A/B)$ even if $[A,B]\neq 0$, as $\tr(A B^{-1})=\tr(B^{-1} A)$.}
\bS\begin{align}
    	\TMU'_{-1,0} &= \left[ \det D_{[2]}(\mu,\mu') \tr\frac{ D_{-1,0}'(\mu,\mu')}{D_{[2]}(\mu,\mu')} \right]_{\mu\mu'} \\
	&= \left[ \tr \frac{\det\big(1 - S_{[2]} (\mu,\mu') \DC_{[2]}\big) \, \DO_{-1,0}}{1 - S_{[2]}(\mu,\mu') \DC_{[2]}} \right]_{\mu\mu'},
\label{eq:Um}\end{align}\eS
where we plugged in $D_{[2]}(\mu,\mu')$ and $D_{-1,0}'(\mu,\mu')$ from \eqref{eq:modDressedGF}, and pulled the determinant inside the trace. 
Inserting Eq.~\eqref{eq:Um} back into Eq.~\eqref{eq:AnaDerivU} allows for an analytic expression of the derivative of the generating functional, and with it for the expectation value.

When placing the measurement somewhere in the center of the Keldysh contour, c.f. Eq.~\eqref{eq:UVGenFunc2}, one would find that
\begin{align} \label{eq:AnaDerivFull}
\expval*{\hat{O}} &= \expval*{\hat{O}}_{-1,0}+\expval*{\hat{O}}_{0,1}-\expval*{\hat{O}}_{0}, 
\end{align}
where $\expval*{\hat{O}}_{-1,0}$ is given by Eq.~\eqref{eq:AnaDerivU}, while $\expval*{\hat{O}}_{0,1} = \lambda_0^{-1} \mel*{\tilde\lambda_0}{\TMU_{0,1}'}{\lambda_0}$ and $\expval*{\hat{O}}_{0} = \mel*{\tilde\lambda_0}{\TMV_{\!0}'}{\lambda_0}$. 
The respective derivatives are then calculated in analogy to Eq.~\eqref{eq:Um}.
 
Using the analytic derivative instead of a numeric derivative reduces numerical errors and of course allows for a more straightforward implementation. 
We now continue to introduce a differential measurement, to decrease the impact of numerical errors. 

\subsubsection{Differential measurement}

In a final step, we minimize discretization errors further by only calculating a differential form of the observable of interest. 
This means, instead of the $U$-dependent observable we only calculate the difference between the interacting case and the noninteracting limit numerically, and get an improved estimate
\begin{align}
    \expval*{\hat{O}} = O^{(0)} + \expval*{\hat{O}(U) - \hat{O}(U=0)}
\end{align}
for the considered observable. 
The analytic expectation values $O^{(0)}$ for $U=0$ are calculated by employing the Meir-Wingreen formula \cite{Meir_1992,Jauho_1994} for the current and the dot's lesser Green's function to account for occupation number and $z$ projection of the spin, leading to
\newcommand{\Aw}{p(\omega)}
\bS\label{eqs:AnaObserv}
\begin{align} 
    I^{(0)} & = 2\,\Gamma^2\int_{-\infty}^\infty \frac{\dd{\omega}}{2\pi} \frac{\Aw\left[ f_\aL(\omega) - f_\aR(\omega) \right]}{[\Aw^2-B^2] (\omega - \epsilon_0)}, \\
    N^{(0)} & = 2\,\Gamma\int_{-\infty}^\infty \frac{\dd{\omega}}{2\pi} \frac{\Aw\left[ f_\aL(\omega) + f_\aR(\omega) \right]}{[\Aw^2-B^2](\omega - \epsilon_0)}, \\
    S_{z}^{(0)} & = B\Gamma \int_{-\infty}^\infty \frac{\dd{\omega}}{2\pi} \frac{f_\aL(\omega) + f_\aR(\omega)}{[\Aw^2-B^2](\omega - \epsilon_0)}, 
\end{align}\eS
where we defined $\Aw=[\Gamma^2 + (\omega - \epsilon_0)^2]/(\omega - \epsilon_0)$.
Since it is expected that the numerical calculation of the observable at $U=0$ has similar errors as in the interacting case, using the differential measurement cancels these errors. 
This leads to a significant reduction of numerical errors. 
The remaining errors are then effectively eliminated during an extrapolation procedure, which is discussed in the next section. 

\subsection{Two-step extrapolation}\label{ssec:extrapolation}
\begin{figure}[t]
\centering
\includegraphics[width=\columnwidth]{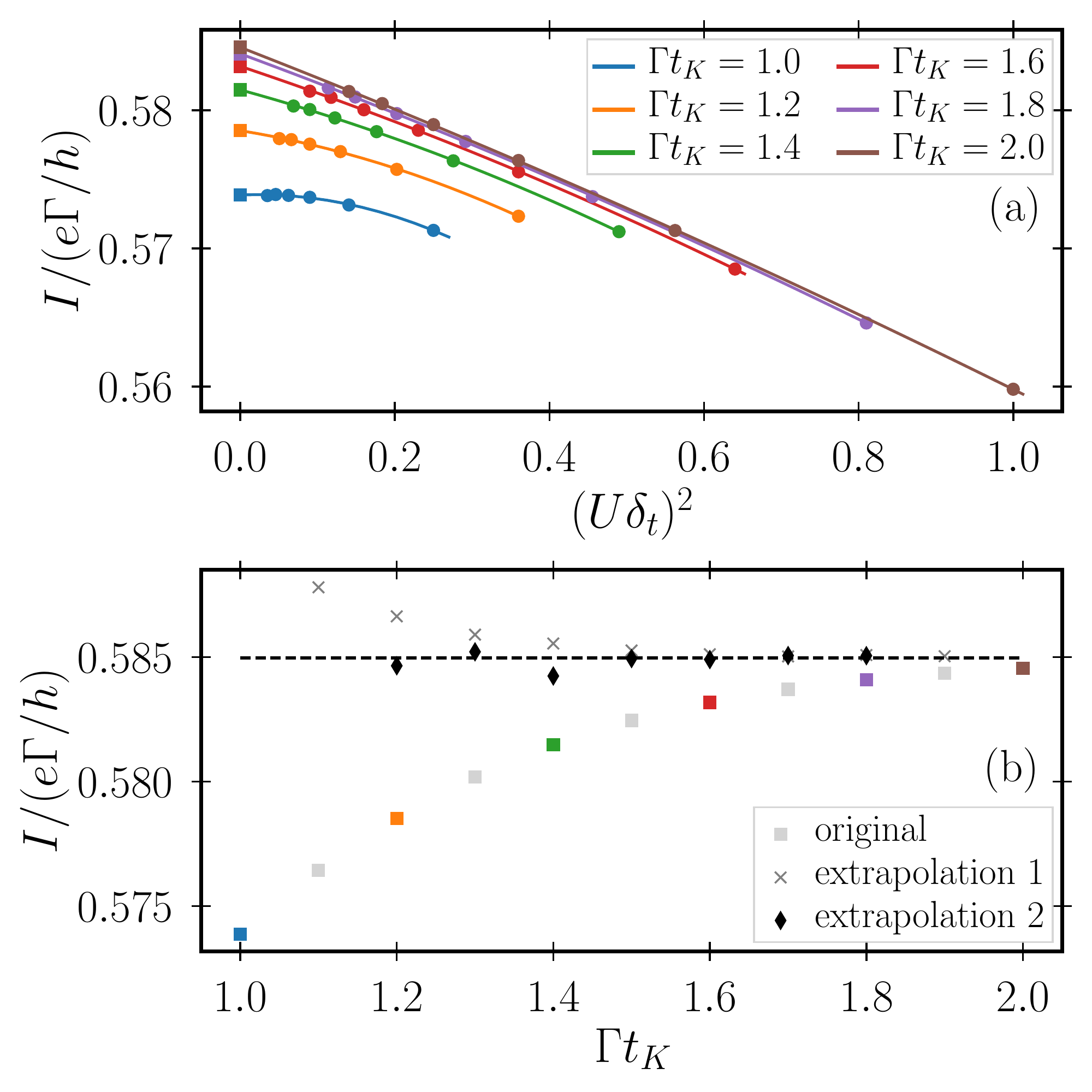}
\caption{The extrapolation procedure used to eliminate systematic errors for the example of the current as an observable. 
(a) Regression of $\delta_t^2 \to 0$ to eliminate the Trotter error, for different $t_K$ with $K=3, \ldots,8$ and $U=1.5\,\Gamma$. 
The squares at $\delta_t^2=0$ are the resulting values of this procedure. 
They can be found again in  the ``original'' data set in (b), where the current is shown as a function of $\Gamma t_K$. 
For this data we employ the Aitken extrapolation \eqref{eq:Aitken} twice, shown as ``extrapolation 1'' and ``extrapolation 2''. 
From the mean of the data of iteration 2 the limiting value $t_K \to \infty$ is received (shown as a dashed line, error estimate is of the order of the line width).}
\label{fig:Converge}
\end{figure}

Throughout the derivation of Eq.~\eqref{eq:AnaDerivU}, we introduced two systematic errors; one being the Trotter error, caused by the finite discretization length $\delta_t$ \cite{Fye_1986}, the other is the error introduced by the truncation at memory time $t_K$. 
However, both can be eliminated using an extrapolation procedure which allows us to provide numerically exact data. 
In earlier works \cite{Weiss_2008, Mundinar_2020}, different approaches were used for this extrapolation procedure, the most common being a two-step regression, first for $\delta_t \to 0$, and then with the resulting values for $1/t_K\to 0$. 
We refer to the aforementioned sources for a detailed discussion of these procedures. 
The main problem of these procedures is that the second regression $1/t_K \to 0$ using a power series ansatz is difficult to motivate. 
Therefore, we employ, in this work, a more sophisticated extrapolation procedure, that also starts with a power series regression of $\delta_t \to 0$ but then uses Aitken extrapolations \cite{Aitken_1927}, which are known to become exact for purely exponential sequences. 

The complete process is shown in Fig.~\ref{fig:Converge} for the example of the current as an observable and for the parameter set $\kB T = 0.5\,\Gamma$, $\epsilon_0 = \Gamma$, $eV = 0.5\,\Gamma$, and $U = 1.5\,\Gamma$.

First, we calculate the expectation value $O$ of the observable $\hat{O}$ for a fixed memory time $t_K$ and varying step size $\delta_t$. 
The result is a set of different realizations of the same observable for this specific parameter set and memory time, $O(t_K, \delta_{t} = t_K/K)$ with $K=3,\ldots,8$. 
It is well known that Trotter errors are of the order $\delta_t^2$ \cite{Fye_1986}. 
Consequently, we fit $O(t_K, \delta_t)$ against a polynomial expression
\begin{align}
    O(t_K, \delta_{t} \to 0 ) = \lim_{\delta_t\to 0} \sum_{j=0}^n c_j \delta_t^{2j}, 
\end{align}
such that the observable with eliminated Trotter error $O(t_K)$ is given by the constant $c_0$ (see Fig.~\ref{fig:Converge}(a)). 
Note that in the equation above it is sufficient to stop at $n=2$, thus only taking up to second order in $\delta^2_t$ into account \cite{Weiss_2008, Mundinar_2020}. 

Having eliminated the Trotter error, we now turn to eliminate the truncation error. 
For this, we repeat the first step for different values of the memory time $t_K$, with $1 \leq \Gamma t_K \leq 2$. 
This leads to a set of realizations of the desired observable $O(t_K)$. 
As can be seen in Fig.~\ref{fig:Converge}(b), the observable as a function of $\Gamma t_K$ converges exponentially against a limiting value for $t_K\to \infty$, which is the numerically exact result of the observable for one specific parameter set. 
This behavior can be understood from the exponential decay of the Green's function $\Delta$, cf.~Fig.~\ref{fig:GFdecay}.
For such exponentially converging sequences the Aitken extrapolation works exceptionally well, accelerating the convergence, which eventually leads to an approximately constant sequence at the limiting value. 
The Aitken extrapolation for a sequence $f_n$ is given by \cite{Aitken_1927}
\begin{align}\label{eq:Aitken}
    (\mathcal A f)_{n+1} = f_n - \frac{(\Delta_n f_n)^2}{\Delta_n^2 f_n},
\end{align}
with forward differences \cite{KoenigHucht21} $\Delta_n f_n=f_{n+1}-f_n$.
For $f_n$ we use the set of realizations of the desired observable $O(t_K)$ and find, that after two Aitken extrapolations the data is approximately constant around the numerically exact result for sufficiently large values of $t_K$.
We use the mean value of this sequence as the final result and its standard deviation as an error estimate. 

\section{Results} \label{sec:Results}

\begin{figure}[t]
\centering
\includegraphics[width=\columnwidth]{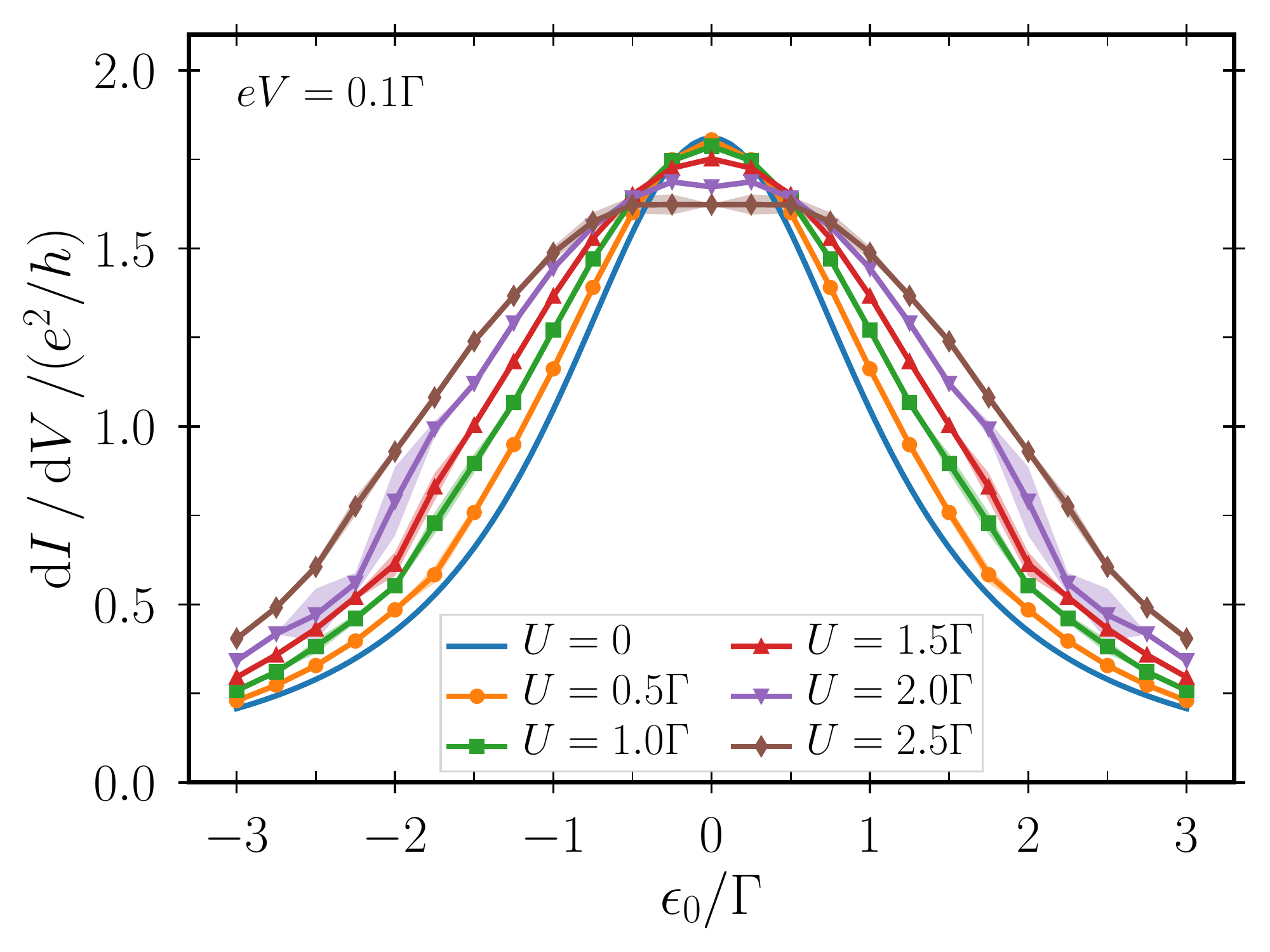}
\caption{The differential conductance as a function of the gate voltage $\epsilon_0$ for different values of the Coulomb interaction strength $U$. 
Shown is the regime of linear response, given by small bias voltages $eV = 0.1\,\Gamma$. 
Other parameters are $\kB T=0.2\,\Gamma$ and $B=0$, shaded areas are error estimates.}
\label{fig:Ieps0DiffU}
\end{figure}

Having introduced and extensively discussed the TraSPI method, we now use it to calculate current-based and occupation number-based observables of a single-level, interacting quantum dot coupled to two normal leads. 
To make interaction-induced effects more visible, we show derivatives of the current and the occupation number with respect to the bias voltage $eV$. 
These derivatives are calculated numerically via central differences, 
\begin{align}
     \dv{O(V)}{V} = \frac{O(V+\delta_V) - O(V-\delta_V)}{2 \delta_V}+\mathcal{O}(\delta_V^2),
\end{align}
where we choose $e\delta_V \leq 0.01\,\Gamma$. 

For all data sets we took memory sizes $1.0 \leq \Gamma t_K \leq 2.5$ and $K = 3,\ldots,7$ into account. Whenever this was not sufficient to reach convergence, we included $K=8$. 

\subsection{Differential conductance}

We start the discussion of the conductance in the linear-response regime with Fig.~\ref{fig:Ieps0DiffU}. 
There, the $\dd I/\dd V$ is shown as a function of the gate voltage $\epsilon_0$ for a small bias voltage $eV=0.1\,\Gamma$. 
The other parameters are given by $\kB T = 0.2\,\Gamma$ and $B=0$. 
Each curve represents a different value of the Coulomb interaction strength $U$, starting at $U=0$, which is calculated analytically, see Eqs.~\eqref{eqs:AnaObserv}. 
For such small bias voltages, we are able to reach Coulomb interaction strengths of $U \leq 2.5\,\Gamma$, with the data sets still converging.
For larger $U$ it would be necessary to take memory times $\Gamma t_K > 2.5$ and thus larger $K$ into account. 
For the noninteracting case, we find a peak at $\epsilon_0=0$, where the single level is within the transport window.
The peak height is, for the chosen parameters, at $\dd I/\dd V \approx 1.81 \,e^2/h$.
The deviation from the maximally possible value of $2 \,e^2/h$ is due to finite temperature.
As $\epsilon_0$ moves away from zero, the dot's energy level is pushed out of the transport window and the conductance drops significantly. 
When increasing the interaction strength $U$, one would expect a level splitting for single and double occupation of the quantum dot, and as a result two peaks at $\epsilon_0 = \pm U/2$ in the $\dd I/\dd V$ curves. 
We do not reach high enough values of $U$ to clearly resolve this peak splitting, but we see that the central peak becomes significantly broader. 
In addition, at $\epsilon_0 = 0$ the conductance is suppressed with increasing $U$, with the maximal value for $U=2.5\,\Gamma$ being $\dd I/\dd V = 1.623 \pm 0.002 \,e^2/h$. 

\begin{figure}[t]
\centering
\includegraphics[width=\columnwidth]{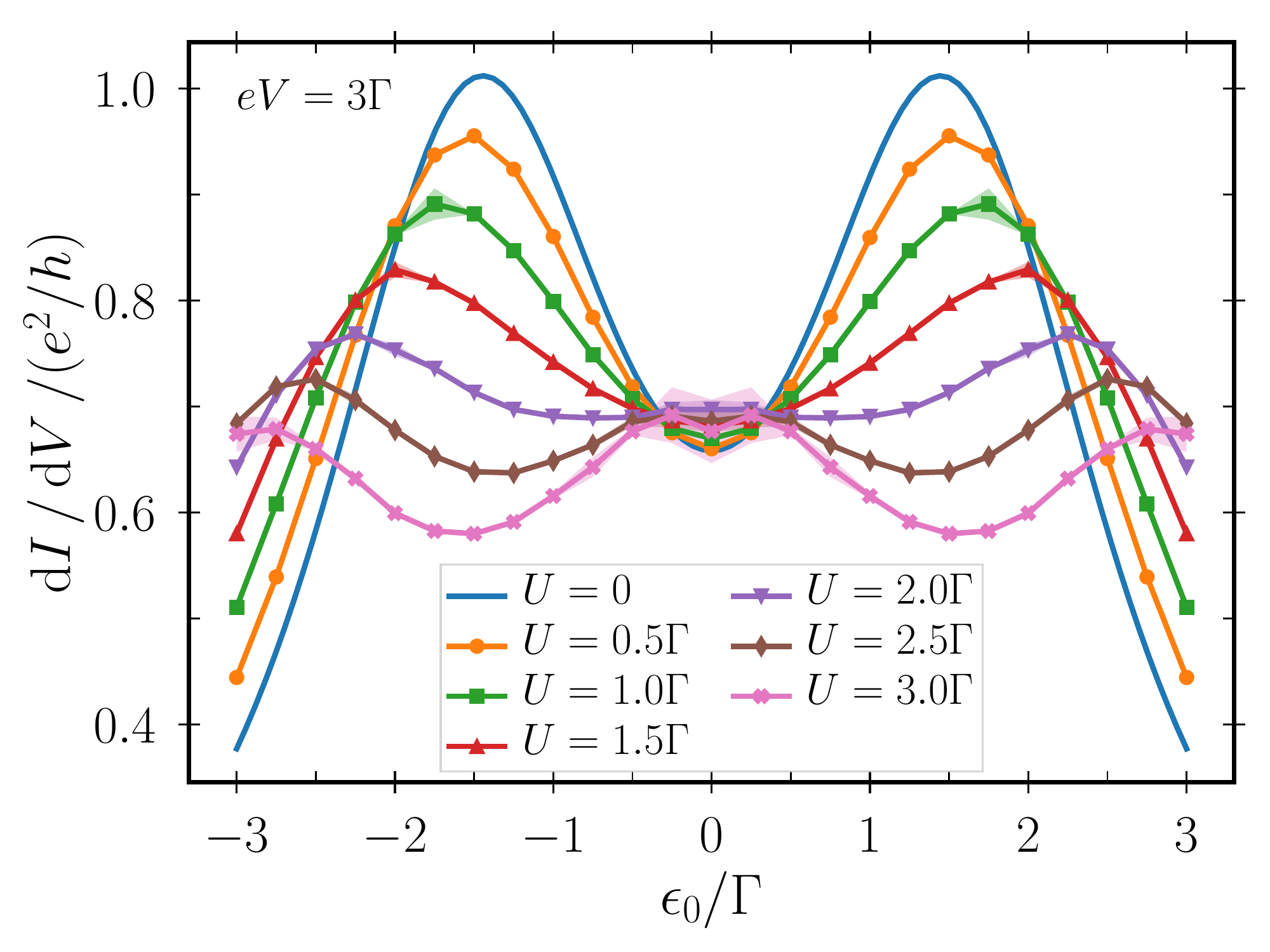}
\caption{The differential conductance as a function of the gate voltage $\epsilon_0$ for different values of the Coulomb interaction strength $U$. 
Parameters are $\kB T=0.2\,\Gamma, eV=3\,\Gamma$, $B=0$.}
\label{fig:dIdVeps0DiffU}
\end{figure}

Next, we turn to the nonlinear-response regime. 
In Fig.~\ref{fig:dIdVeps0DiffU}, we show the differential conductance for a large bias voltage of $eV = 3\,\Gamma$. 
The other parameters are as in Fig.~\ref{fig:Ieps0DiffU}.
Due to the large bias voltage, the $\epsilon_0$ dependence of the conductance is more complex. 
For $U=0$, there are two peaks at $\epsilon_0 \approx \pm eV/2 = \pm 1.5\,\Gamma $, reflecting the two resonance conditions of the dot level matching the Fermi level of the left and right electrode, respectively.
The peaks are not perfectly centered around $\pm eV/2$ due to the finite width of the peaks. 
Since the two peaks overlap, their maxima get pulled towards each other.

At finite Coulomb interaction, there are, in principle, four resonance conditions, given by $\epsilon_0 \approx \pm eV/2 \pm U/2$.
This explains that with increasing the Coulomb interaction $U$, the two peaks are pushed away from each other.
Simultaneously, the peak heights decrease significantly.
This is a consequence of the reduced overlap of the peaks as they move away from each other.

The most remarkable feature, however, is the appearance of a third peak around $\epsilon_0=0$ for $U > 2\,\Gamma$.
This is due to the fact that here $U \approx eV$, and therefore at $\epsilon_0 = 0$ the singly occupied state is in resonance with the right lead and the doubly occupied state is in resonance with the left lead. 
Since a noninteracting-electron picture only predicts two peaks, the appearance of additional peaks is clear indication of Coulomb interaction. 
However, a third (and ultimately a fourth) peak can only be resolved for sufficiently large values of $U$.
This, we could not achieve by using the finite-time implementation of ISPI, but the TraSPI formulation now allows us to enter this regime.

\begin{figure}[t]
\centering
\includegraphics[width=\columnwidth]{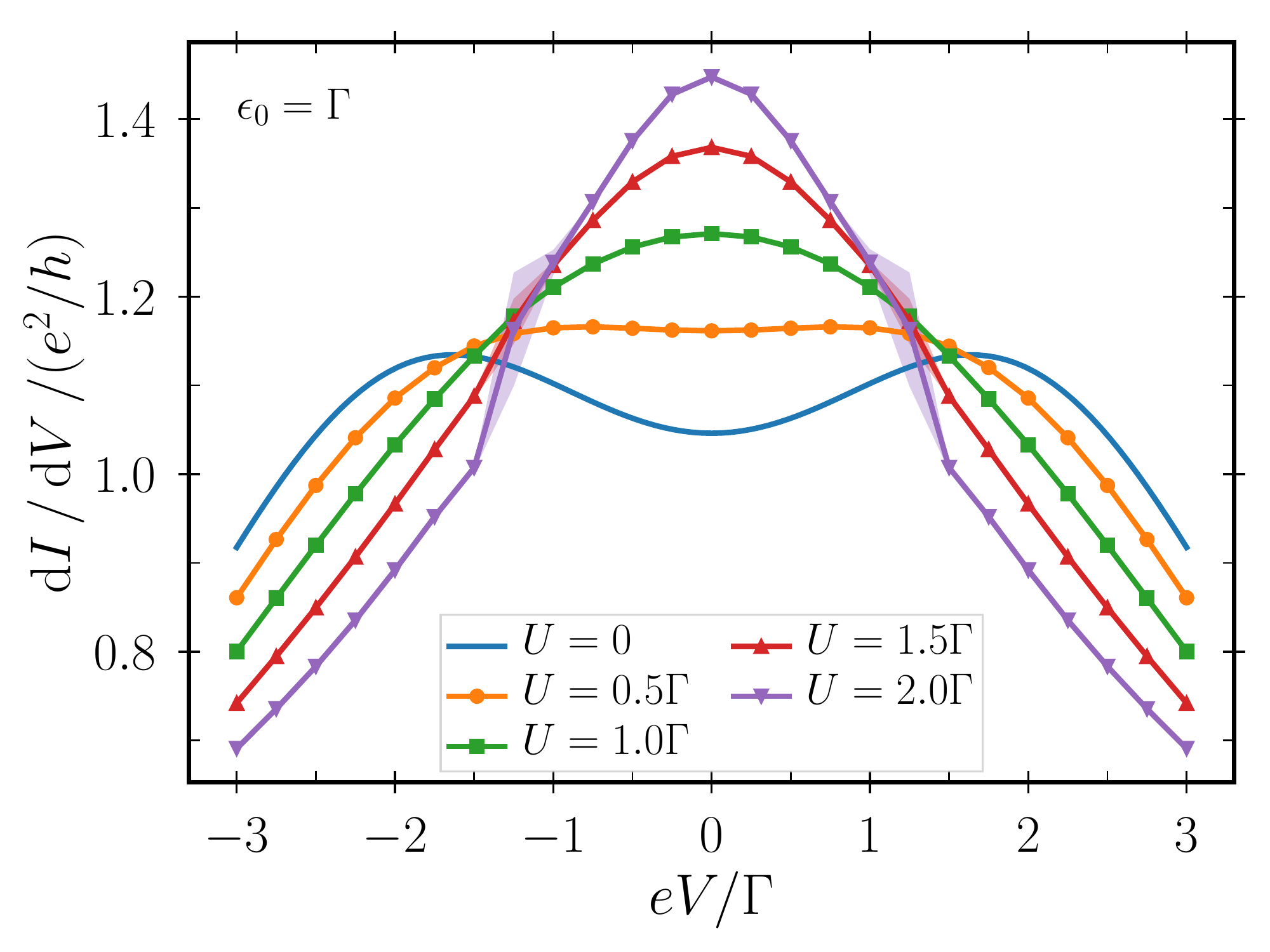}
\caption{The differential conductance as a function of the bias voltage $eV$ for different values of the Coulomb interaction strength. 
Parameters are $\kB T=0.2\,\Gamma$, $\epsilon_0=\Gamma$, $B=0$.}
\label{fig:dIdVeVDiffU}
\end{figure}

To address the crossover from the linear- to the nonlinear-response regime, we discuss in Fig.~\ref{fig:dIdVeVDiffU} the differential conductance as a function of the bias voltage at $\epsilon_0 = \Gamma$. 
The temperature is again $\kB T = 0.2\,\Gamma$ and $B=0$.
For vanishing Coulomb interaction, we find, as expected, two peaks located at $eV \approx \pm \epsilon_0 =\pm \Gamma$. 
With increasing Coulomb interaction $U$, the differential conductance increases in the linear-response regime but decreases in the nonlinear-response regime, in accordance to our findings in Figs.~\ref{fig:Ieps0DiffU} and \ref{fig:dIdVeps0DiffU}. 

We emphasize that the system under consideration is particle-hole symmetric, such that the differential conductance $\dd I/\dd V$ is an even function of both $\epsilon_0$ and $eV$. 
Both symmetries are fulfilled perfectly by the TraSPI formalism up to numerical accuracy. 

\subsection{Occupation number}

\begin{figure}[t]
\centering
\includegraphics[width=\columnwidth]{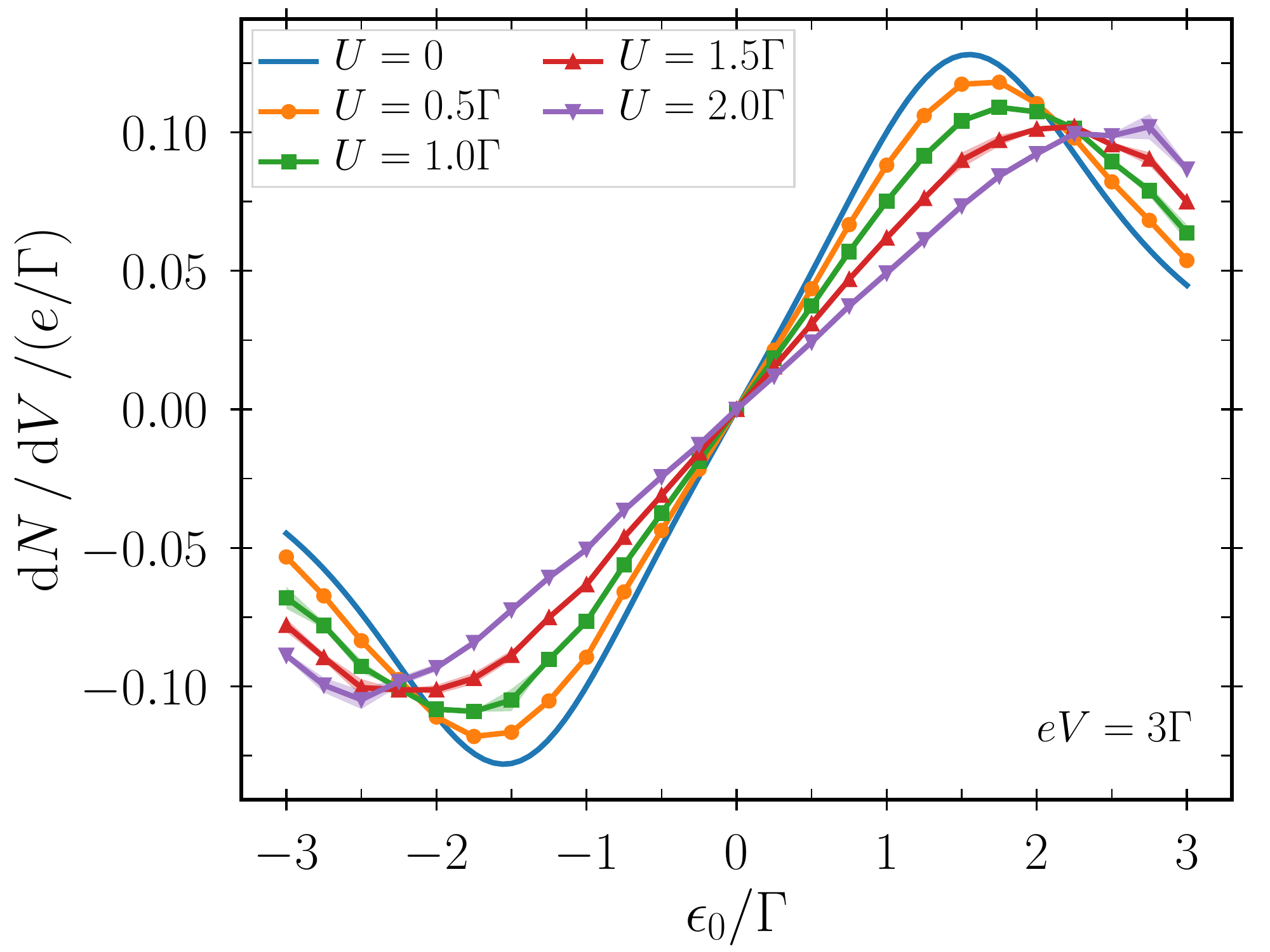}
\caption{The derivative of the occupation number with respect to bias voltage in the nonlinear-response regime, $eV=3\,\Gamma$, as a function of gate voltage $\epsilon_0$. Shown are different values of the Coulomb interaction strength $U$. 
Parameters are $\kB T=0.2\,\Gamma$, $B=0$.}
\label{fig:dNdVeps0higheV}
\end{figure}

We now address the average number of electrons on the quantum dot.
For low-lying $\epsilon_0$, the quantum dot is occupied with two electrons.
With increasing $\epsilon_0$, the occupation number is reduced and ultimately becomes empty.

In Fig.~\ref{fig:dNdVeps0higheV}, we show the derivative of the occupation number as a function of gate voltage in the nonlinear-response regime.
We choose the same parameters as in Fig.~\ref{fig:dIdVeps0DiffU}.
Similarly as the differential conductance is better suited to resolve detailed structures, the derivative of the occupation number shows more structure than the occupation number itself.
In that respect, Fig.~\ref{fig:dIdVeps0DiffU} for the current and Fig.~\ref{fig:dNdVeps0higheV} for the occupation number are analogous to each other.
The peaks of $\dd N/\dd V$ are at the same positions as the peaks of $\dd I/\dd V$, indicating that a large change of the current is accompanied with a large change of the occupation number.
With increasing $U$, the peaks move away from each other, and the TraSPI formalism is able to accurately describe the full region in between.

The numerical calculations of the conductance and the occupation number are independent of each other.
The information contained in these two quantities is partially the same but in most transport regimes differs from each other in detail.
There is, however, one limit in which they carry identical information.
This occurs at zero temperature, $T=0$, and vanishing bias voltage $V=0$.
In this case, the electrons transversing the quantum dot scatter only elastically, such that Friedel's sum rule can be applied, which leads to the Langreth formula \cite{Langreth_1966, Ng_1988}
\begin{align} \label{eq:sumrule}
    \frac{\dd I}{\dd V}\bigg|_{T,V=0}=\frac{2e^2}{h} \frac{4\,\Gamma_\aL \Gamma_\aR}{(\Gamma_\aL+\Gamma_\aR)^2} \sin^2 \left( \frac{\pi}{2} \big\langle N\big\rangle \right).
\end{align}
This remarkable result is valid for any value of the interaction strength $U$, covering all regimes from noninteracting electrons to strong correlations, e.g., in the Kondo regime.
The Langreth formula is, in general, violated for any approximation scheme that only includes a certain class of transport contributions.
However, for a numerically exact treatment, which includes TraSPI, the Langreth formula serves as a consistency check to assess the quality of the method.

\begin{figure}[t]
\centering
\includegraphics[width=\columnwidth]{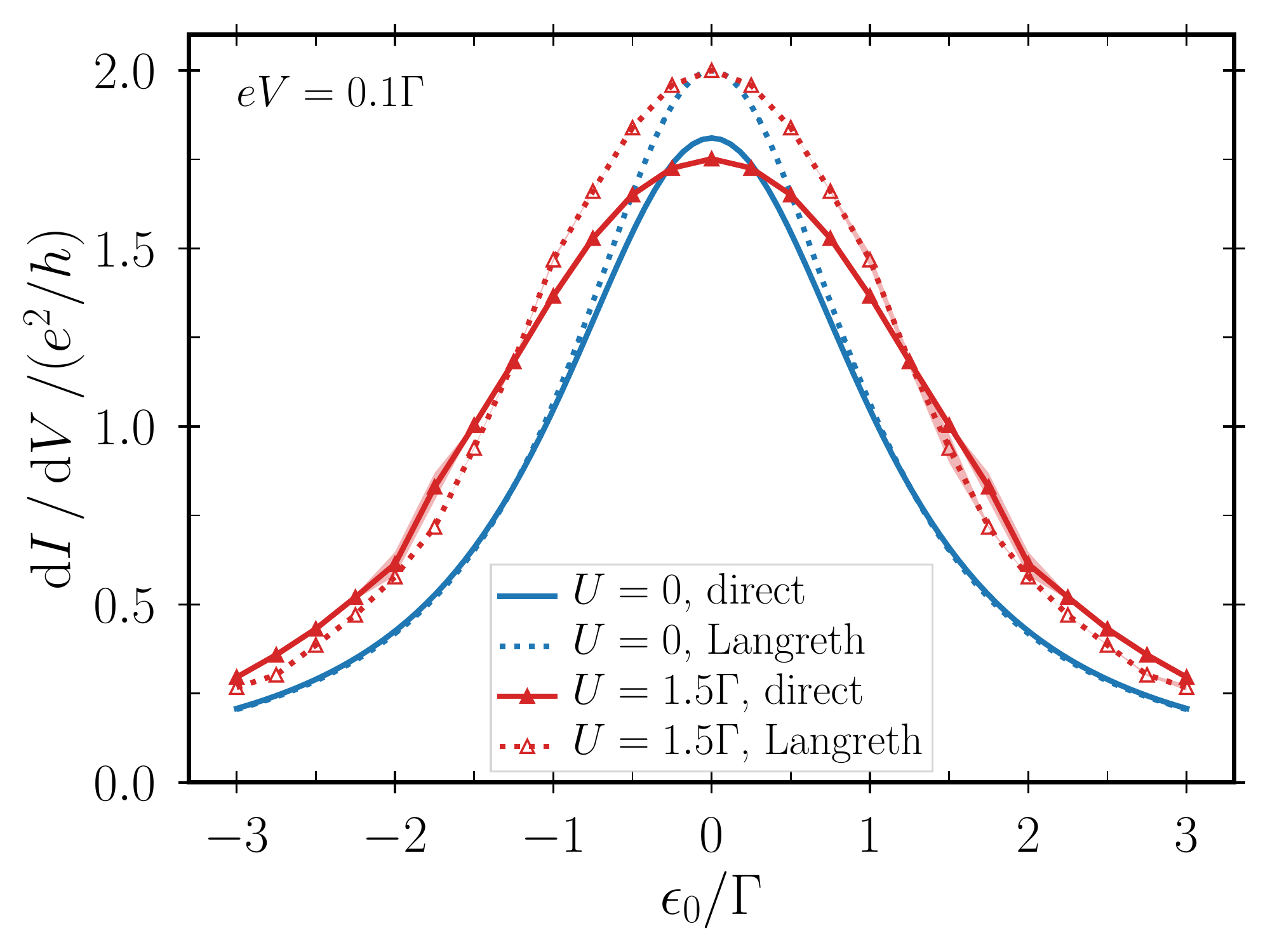}
\caption{The differential conductance for low bias voltage $eV=0.1\,\Gamma$, as a function of the gate voltage $\epsilon_0$ for different values of the Coulomb interaction strength $U$. 
Comparison between the direct result obtained via the TraSPI scheme (direct) and the result obtained from Langreth's formula \eqref{eq:sumrule}.
Parameters are $\kB T=0.2\,\Gamma$, $B=0$.}
\label{fig:GOccComparison}
\end{figure}

In Fig.~\ref{fig:GOccComparison}, we compare the linear conductance calculated in two different ways: once directly and once via the calculation of the occupation number and Langreth's formula Eq.~\eqref{eq:sumrule}.
As stated above, Langreth's formula holds at zero temperature. 
This regime can be accessed by TraSPI only away from resonance, whereas close to the resonance condition $\epsilon_0=0$, numerical convergence is too slow.
Therefore, we perform our calculations at finite temperature, choosing $\kB T = 0.2\,\Gamma$, in accordance with the previous figures.

We find very good agreement away from resonance, where the influence of finite temperature on the conductance is small. 
The deviation of the Langreth result Eq.~\eqref{eq:sumrule} from the direct calculation at resonance $\epsilon_0$ is fully understood as a finite-temperature effect. 
In conclusion, Fig.~\ref{fig:GOccComparison} gives us strong confidence in the quality of TraSPI as a numerically exact method.

\section{Conclusions} \label{sec:Conclusion}

In the literature, there is a plethora of methods to describe quantum transport through nanostructures, which all have their advantages and disadvantages in different regimes.
Some are restricted to the linear-response regime while others can cover strong nonequilibrium situations.
In scenarios with a clear hierarchy of the involved parameters, a perturbation expansion in one of them can be used.
However, in real experiments very often many of these parameters characterizing, e.g., temperature, tunnel coupling strength, Coulomb interaction as well as gate and bias voltage, are of the same order of magnitude.
Then, numerically exact methods are desirable.

In this paper, we presented TraSPI as such a numerically exact method.
It is based on an iterative summation of path integrals, referred to as ISPI.
The virtue of ISPI (and thus also TraSPI) is that it naturally takes into account all orders in tunneling of electrons between quantum dot and leads, allows for arbitrary bias voltages that drive the system out of equilibrium, is not restricted to either low or high temperature, and is able to include finite Coulomb interaction.
While the method is numerically exact after a suitable extrapolation procedure, stronger correlations increase the convergence time, such that ISPI (and TraSPI) are best suited for small to intermediate strengths of the Coulomb interaction.
In previous applications of the ISPI method \cite{Mundinar_2019,Mundinar_2020}, we concentrated on spin-dependent phenomena, which show an interaction dependence already for moderate Coulomb interaction strengths.
To increase the range of applicability towards stronger Coulomb interaction strengths, the efficiency of the method needs to be increased.

This we do in the present paper by mapping the ISPI scheme to a transfer-matrix approach, which results in TraSPI.
The major virtue of involving transfer matrices is that the stationary limit is implemented by construction.
This avoids the numerically costly extrapolation of the results of the finite-time formulation as done in ISPI.
In addition, the use of transfer matrices allows for further improvements that enhance the efficiency of the method.
Numerical effort is reduced by optimally choosing the position of the measurement in time. 
Furthermore, to minimize numerical errors, it is advantageous to analytically implement derivatives, e.g., of the current with respect to bias voltage to get the differential conductance, instead of performing a numerical derivative of the current.
And finally, we reduce numerical errors by making use of the possibility to calculate the noninteracting limit analytically and to numerically calculate only the difference between the interacting and the noninteracting case. 

To illustrate the performance of the TraSPI method, we analyzed the differential conductance through a single-level quantum dot in both the linear and nonlinear regime.
We were able to reach values of the Coulomb-interaction strength that are sufficient to resolve in the non-linear response regime a third conductance peak instead of only two peaks that are expected for non-interacting electrons.
Finally, we were able to perform a quality check of TraSPI by demonstrating that Langreth's formula, which connects the zero-temperature, linear conductance with the dot's occupation, is fulfilled in the regime of its applicability range.
Therefore, we are confident that the TraSPI formulation enables us to address systems, transport regimes and effects that could not be covered by previous methods.

\section{Acknowledgements}
We thank S.~Weiss for fruitful discussions on the ISPI scheme. 
Financial funding of the Deutsche Forschungsgemeinschaft (DFG, German Research Foundation) under project 278162697 -- SFB 1242 is acknowledged.

\end{document}